\documentstyle[preprint,aps,eqsecnum]{revtex}

\begin{document}
{\preprint{PITT-96-7777.7;  CMU-HEP-96-11;  
LPTHE-96-19; DOR-ER/40682-122}
\draft
\title{{\bf OUT OF EQUILIBRIUM DYNAMICS OF AN INFLATIONARY
 PHASE TRANSITION }}
\author{{\bf D. Boyanovsky$^{(a)}$, D. Cormier$^{(b)}$, H. J. de Vega$^{(c)}$,
and R. Holman$^{(b)}$}}   
\address
{ (a)  Department of Physics and Astronomy, University of
Pittsburgh, Pittsburgh, PA. 15260, U.S.A. \\
 (b) Department of Physics, Carnegie Mellon University, Pittsburgh,
PA. 15213, U. S. A. \\ 
 (c) Laboratoire de Physique Th\'{e}orique et Hautes Energies, Universit\'{e}
Pierre et Marie Curie (Paris VI), Tour 16, 1er. \'{e}tage, 4, Place Jussieu
75252 Paris cedex 05, France}
\date{October 1996}
\maketitle
\begin{abstract}

We study the non-linear dynamics of an inflationary phase transition in a
quartically self coupled inflaton model within the framework of a de Sitter
background.  Large $ N $ and Hartree non-perturbative approximations combined with
non-equilibrium field theory methods are used to study the self-consistent time
evolution including backreaction effects. We find that when the system cools
down from an initial temperature $T_i > T_c$ to below $T_c$ with the initial
value of the zero mode of the inflaton $\phi(0) << m \lambda^{-1/4}$, the
dynamics is determined by the growth of long-wavelength quantum
fluctuations. For $\phi(0) >> m \lambda^{-1/4}$ the dynamics is determined by
the evolution of the classical zero mode. In the regime where spinodal quantum
fluctuations give the most important contribution to the non-equilibrium
dynamics, we find that they modify the equation of state providing
a graceful exit from the inflationary stage.  Inflation ends through this new
mechanism at a time scale $t_s \geq [H / m^2]\ln \left[ \lambda^{-1} \right] $
which for $H \geq m$ and very weak coupling allows over one hundred e-folds
during the de Sitter phase.  Spatially correlated domains grow to be of
horizon-size and quantum fluctuations ``freeze-out'' for times $t> t_s$.

\end{abstract}

\section{\bf Introduction}

For all of its observational success, the standard big bang cosmology is not
without its shortcomings.  The inflationary scenario was conceived\cite{guth}
to overcome many of these shortcomings, chief amongst them the flatness and
horizon problems.  The original old inflationary scenario \cite{guth} required
a first-order phase transition with strong supercooling, during which the scale
factor grew exponentially thus solving the homogeneity and flatness
problem. This model, however, lacked any mechanism to provide a succesful exit
from the inflationary phase\cite{problem}.  The new inflationary or
`slow-rollover' scenario \cite{linde1,andy} overcame these problems. In these
scenarios, a phase transition from an unbroken phase to a symmetry broken one
takes place, and the scalar field, whose expectation value serves as the order
parameter, drives inflation. This scalar field, the inflaton, drives a stage of
exponential (de Sitter) expansion while it slowly rolls down the potential
hill.  Other versions, such as chaotic inflation\cite{chaotic} in which a phase
transition is not necessary have been advocated (for reviews of inflationary
cosmology, see\cite{rev1,rev2,kolbook,lindebook}).

Phase transitions play an important role in the early universe, and, as
mentioned above, are pivotal to many inflationary models. They are also
necessary in models in which topological defects provide the seeds for density
perturbations\cite{topo}.  It has been pointed out by many authors that using
flat space-time effective potentials and equilibrium concepts to describe the
dynamics of the inflationary phase transition can be misleading. Both
gravitational effects as well as those due to quantum and thermal fluctuations
have been studied and found to be very important for the description of the
dynamics\cite{vil,linde2,vilford,vilenkin1,mazenko,guthpi}.

Although there have been previous studies of the dynamics of phase
transitions\cite{vilenkin1,mazenko,guthpi}, most of the attempts were hampered
by the lack of a self-consistent non-perturbative treatment of the
non-linearities and backreaction effects.  It is only recently that
non-equilibrium methods have been developed to study real time phenomena during
phase transitions\cite{hu,ringwald,leutwyler,weiss} and other strongly out of
equilibrium situations\cite{motolla}, including non-perturbative
treatments\cite{largen,jackiwetal,guven}.  During the last few years these
methods have been adapted to study strongly out of equilibrium phenomena in
cosmology, allowing the understanding of the dynamics of non-linear field
theory in such situations\cite{frw,boyrehe}.

Studies of the dynamics of typical second order phase transitions in Minkowski
space-time have revealed a wealth of new and interesting phenomena. In
situations in which the phase transition occurs on time scales which are faster
than the relaxation time of long-wavelength fluctuations, the dynamics is
driven by the growth of long-wavelength, spinodal instabilities.  The
backreaction of these quantum and thermal fluctuations dramatically modifies
the evolution of the zero mode of the scalar field eventually shutting off the
instabilities \cite{roller,domains,cooper}.  These instabilities are the
hallmark of the phase transition and are responsible for the formation and
growth of correlated domains\cite{domains,cooper}.

The nature of these instabilities and their intrinsic non-perturbative dynamics
is rather simple to understand\cite{roller,domains,cooper}. As the system cools
down from a quasi-equilibrium disordered high temperature phase to below the
critical temperature the effective mass of the scalar field becomes
negative. If the expectation value of the field was zero in the equilibrium
high temperature phase, it will remain zero below the critical temperature in
the absence of biasing fields.  The fluctuations of the field will grow in time
and eventually become large enough to sample the minima of the (effective)
potential, which are non-perturbatively large in amplitude. This growth of
fluctuations translates into a growth of spatial correlations of the field,
resulting in domains\cite{domains}. In Minkowski space-time there is no
restriction to the final size of these domains, and generically in weakly
coupled theories these correlated regions reach sizes several times larger than
the zero temperature correlation length\cite{domains}. In a de Sitter spacetime
or in a general FRW cosmology, for that matter, causality limits the physical
distance within which the order parameter can be correlated; in a de Sitter
spacetime, this distance is $d_H= H^{-1}$, with $H$ the Hubble constant.

The goal of this article is to study the full non-equilibrium dynamics of an
inflationary phase transition including the backreaction effects in a simple
model of a scalar field (the inflaton) with a quartic self-interaction.  We use
the methods of non-equilibrium field theory\cite{hu} combined with
self-consistent, non-perturbative Hartree and large $N$
approximations\cite{motolla,largen,frw,boyrehe,cooper,largenfrw}.

The metric is taken to be a fixed cosmological background within which we study
the effects of quantum and thermal fluctuations on the evolution of the
expectation value of the inflaton and the energy momentum tensor, both
analytically and numerically, so as to follow the equation of state.
When the temperature falls below the critical value, long-wavelength
fluctuations grow exponentially as a result of the spinodal instabilities and
their contribution to the equations of motion eventually become of the same
order as the tree-level terms.  We argue that in the very weakly coupled case,
$\lambda \approx 10^{-12}$, the change in temperature and the effective mass
occur on time scales much shorter than the time in which long-wavelength
fluctuations can adjust to local thermodynamic equilibrium, and they fall out
of equilibrium just as in a ``quench'' from the high temperature phase.
 
When these fluctuations are incorporated self-consistently as a backreaction in
the evolution equations, we find, for slow-roll initial conditions, that the
effects of these fluctuations dramatically change the dynamics.  The unstable
quantum fluctuations modify the equation of state away from the vacuum
dominated state required for de Sitter expansion, and for a reasonably wide
range of parameters and initial conditions, these fluctuations provide a new
mechanism to gracefully exit the inflationary stage within an acceptable number
of e-folds.  During this time, spatial correlations grow to reach horizon-size
and freeze-out at larger times.

In the large $ N $ approximation, we find that the late time dynamics is
summarized by a very useful sum rule relating the infinite time limit of the
order parameter $ \phi(\infty) $ and the infinite time limit of the quantum
fluctuations $ \langle \psi^2(\infty) \rangle $ as follows,
$$
-|m^2|+\phi^2(\infty) +  \langle \psi^2(\infty) \rangle = 0 \; .
$$
This result implies that the excitations are massless and minimally
coupled. However we find the new result that despite the fact that
the relevant degrees of freedom are massless and minimally
coupled asymptotically,
 their equal time two-point function saturates to a constant
value rather than growing linearly in time. 

In section II we introduce the model, the approximations, and analyze the
subtle but important issues of renormalization. Section III is devoted to
presenting the renormalized equations of motion and the correlation
functions. Section IV and V provide an analytic and numerical study of
different cases, as well as an estimate of the backreaction of the quantum
fluctuations onto the scale factor. Section VI presents a discussion of the
results and the final conclusions and further avenues of
study.

\section{\bf The Model and the Approximations}

The need for a non-perturbative treatment of nonequilibrium quantum field
dynamics stems from the fact that when the temperature falls below the critical
point, the effective mass term becomes negative. This results in an instability
of the long-wavelength modes which grow exponentially for early times after
the transition.  The fluctuations of the scalar order parameter will grow in
time to sample the broken symmetry vacua, leading to large amplitude
fluctuations that can only be treated within a non-perturbative scheme.

There are two approximation schemes that have been used to study the
non-equilibrium dynamics during phase transitions, each with its own advantages
and disadvantages.  The Hartree
factorization\cite{vilenkin1,jackiwetal,boyrehe,domains} has the advantage that
it can treat the dynamics of a scalar order parameter with discrete symmetry,
while its disadvantage is that it is difficult to implement consistently beyond
the lowest (mean field) level.  The advantage of the large $ N $
approximation\cite{motolla,largen,frw,boyrehe,largenfrw} is that it allows a
consistent expansion in a small parameter ($1/N$) and correctly treats
continuous symmetries in the sense that it implements Goldstone's theorem.
Moreover, the large $ N $ expansion becomes a $ \lambda/N $ expansion for small
values of $ \lambda $. Therefore, it may be a reliable approximation for the
typical values of $ \lambda $ used for inflation even when $ N = 1 $. It should
be noted that when spontaneous symmetry breaking is present, the large $ N $
limit always produces massless Goldstone bosons.

Both methods implement a resummation of a select set of diagrams to all orders
and lead to a system of equations that is energy conserving in Minkowski space
time, and as will be shown below, satisfies covariant conservation of the
energy momentum tensor in FRW cosmologies.  Furthermore, both methods are
renormalizable and numerically implementable.  Given that both methods have
advantages and disadvantages and that choosing a particular scheme will
undoubtedly lead to criticism and questions about their reliability, we will
study {\em both}, comparing the results to obtain some universal features of
the dynamics.

We restrict our study to a spatially flat Friedmann-Robertson-Walker universe
with scale factor, $a(t)$, and metric,
\begin{equation}
ds^2 = dt^2-a^2(t)\; d\vec{x}^2,
\end{equation}
and focus in particular on the case of a de Sitter space-time with scale
factor,
\begin{equation}
a(t)=e^{Ht}. \label{desitmetric}
\end{equation}

The action and Lagrangian density are given by,
\begin{equation}
S  =  \int d^4x\; {\cal L},\label{action}
\end{equation}
\begin{equation}
{\cal L}  =   a^3(t)\left[\frac{1}{2}\dot{\vec{\Phi}}^2(x)-\frac{1}{2}
\frac{(\vec{\nabla}\vec{\Phi}(x))^2}{a^2(t)}-V(\vec{\Phi}(x))\right],
\label{lagrangian} 
\end{equation}
\begin{equation}
V(\vec{\Phi})  =  \frac{\lambda}{8N}\left(\vec{\Phi}^2+\frac{2N
M^2}{\lambda}\right)^2 
\; \; ; \; \; 
M^2  = - m^2+\xi\;{\cal R}, \label{potential}
\end{equation}
\begin{equation}
{\cal R}  =  6\left(\frac{\ddot{a}(t)}{a(t)}+
\frac{\dot{a}^2(t)}{a^2(t)}\right), \label{ricciscalar}
\end{equation}
where we have included the coupling of $\Phi(x)$ to the scalar curvature ${\cal
R}(t)$ since it will arise as a consequence of renormalization.  In the de
Sitter universe (\ref{desitmetric}), $ {\cal R} = 12H^2 $.  The canonical
momentum conjugate to $\Phi(x)$ is,
\begin{equation}
\vec{\Pi}(x) = a^3(t)\; \dot{\vec{\Phi}}(x), 
\end{equation}
and the {\it time dependent} Hamiltonian is given by,
\begin{equation}
H(t) = \int d^3x\left\{
\frac{\vec{\Pi}^2(x)}{2a^3(t)}+\frac{a(t)}{2}\;(\nabla\vec{\Phi}(x))^2+
a^3(t)\;V(\vec{\Phi})\right\}. 
\end{equation}

\subsection{\bf The Hartree Approximation} 
To implement the Hartree approximation, we set $N=1$ and write,
\begin{equation}
\Phi(\vec x,t) = \phi(t)+\psi(\vec x, t), \label{zeromode}
\end{equation}
with, 
\begin{equation}
\phi(t) = \langle \Phi(\vec x, t) \rangle \; ; \; \langle \psi(\vec x, t)
\rangle = 0, \label{expec}
\end{equation}
where the expectation value is defined by the non-equilibrium density matrix
specified below, and we have assumed spatial translational invariance,
compatible with a spatially flat metric.  The Hartree approximation is obtained
after the factorization,
\begin{equation}
\psi^3(\vec x,t) \rightarrow 3 \langle \psi^2(\vec x,t) \rangle \psi(\vec x,t),
\label{hart3}
\end{equation}
\begin{equation}
\psi^4(\vec x,t) \rightarrow 6 \langle \psi^2(\vec x,t) \rangle \psi^2(\vec
x,t)- 3 \langle \psi^2(\vec x,t) \rangle^2, \label{hart4}
\end{equation}
where by translational invariance, the expectation values only depend on time.
In this approximation, the Hamiltonian becomes quadratic at the expense of a
self-consistent condition.

At this stage we must specify the non-equilibrium state in which we compute the
expectation values above.  In non-equilibrium field theory, the important
ingredient is the time evolution of the density matrix $\rho(t)$ (see\cite{hu}
and references therein).  This density matrix obeys the quantum Liouville
equation whose solution only requires an initial condition
$\rho(t_i)$\cite{hu,leutwyler,weiss,frw,boyrehe}.  The choice of initial
conditions for this density matrix is an issue that pervades any calculation in
cosmology. Since we want to study the dynamics of the phase transition, it is
natural to consider initial conditions that describe the {\em instantaneous}
modes of the time dependent Hamiltonian as being initially in local
thermodynamic equilibrium at some temperature $T_i> T_c$. Given this initial
density matrix, we then evolve it in time using the time dependent Hamiltonian
as in\cite{frw} or alternatively using the complex time path integral method as
described in\cite{hu,leutwyler,weiss,motolla,boyrehe}.

Following the steps of references\cite{boyrehe,roller,domains} we find the
equation of motion for the expectation value of the inflaton field to be,
\begin{equation}
\ddot{\phi}(t)+3H\dot{\phi}(t)+M^2\phi(t)+\frac{\lambda}{2}\phi^3(t)+
\frac{3\lambda}{2}
\phi(t)\langle \psi^2(t)\rangle=0. \label{hartzeromode}
\end{equation}
The equal time correlation function is obtained from the coincidence
limit of the non-equilibrium
Green's functions, which are obtained from the mode functions obeying
\begin{equation}
\left[\frac{d^2}{dt^2}+3H\frac{d}{dt}+\omega^2_k(t) \right]f_k(t)= 0,
\label{hartmodes}
\end{equation}
with the effective  frequencies,
\begin{equation}
\omega^2_k(t) =\frac{k^2}{a^2(t)}+ M^2(t) \; , 
\label{hartfreq}
\end{equation}
where
\begin{equation}
M^2(t) =  M^2+ \frac{3\lambda}{2}\phi^2(t)+
\frac{3\lambda}{2}\langle \psi^2(t) \rangle \; .
\label{hartmass}
\end{equation}

At this stage we must provide the initial conditions on the mode
functions $f_k(t)$. As
mentioned above our choice of initial conditions on the density matrix
is that of
local thermodynamic equilibrium for the instantaneous modes of the time
dependent
Hamiltonian at the initial time. Therefore we choose the initial
conditions on the mode
functions to represent positive energy particle states of the
instantaneous Hamiltonian
at $t=0$, which we chose as the initial time. Therefore our choice of boundary
conditions at $t=0$ is
\begin{equation}
f_k(0)= \frac{1}{\sqrt{W_k}} \; ; \; \dot{f}_k(0) = -i \sqrt{W_k} \; \; ; \; \;
W_k= \sqrt{k^2+M^2_0}, \label{hartbc}
\end{equation}
 where the mass $M_0$ determines the frequencies $\omega_k(0)$ and will be
obtained explicitly later. With these boundary conditions, the mode functions
$f_k(0)$ correspond to positive frequency modes (particles) of the
instantaneous quadratic Hamiltonian for oscillators of mass $M_0$.

The equal time correlation function is given in terms of the mode functions
as\cite{motolla,frw,roller,domains,cooper},
\begin{equation}
\langle \psi^2(t) \rangle = \int
\frac{d^3k}{(2\pi)^3}\frac{|f_k(t)|^2}{2}\coth\left[\frac{W_k}{2T_i}\right].
\label{hartfluc}
\end{equation}
The energy and pressure density are given by\cite{boyrehe},
\begin{eqnarray}
\varepsilon = & & \frac{1}{2}\dot{\phi}^2(t)+
\frac{\lambda}{8}\left(\phi^2(t)+\frac{2M^2}{\lambda}\right)^2 + \nonumber \\
&& \frac{1}{2} \int \frac{d^3k}{2(2\pi)^3 }\coth\left[\frac{W_k}{2T_i}\right]
\left[ |\dot{f}_k(t)|^2+ \omega^2_k(t)|f_k(t)|^2\right] -
\frac{3\lambda}{8}\langle \psi^2(t) \rangle^2,
\label{hartenergy} 
\end{eqnarray}

\begin{equation}
p+\varepsilon = \dot{\phi}^2(t) + \int \frac{d^3k}{2(2\pi)^3
}\coth\left[\frac{W_k}{2T_i}\right] \left[ |\dot{f}_k(t)|^2+
\frac{k^2}{3a^2(t)}|f_k(t)|^2\right]. \label{pplusehart}
\end{equation}

It is straightforward to show using the equations of motion
(\ref{hartzeromode},\ref{hartmodes}) that the bare energy is covariantly
conserved,
\begin{equation}
\dot{\varepsilon}+3H(p+\varepsilon)=0. \label{hartcons}
\end{equation}

\subsection{\bf The Large $N$ Approximation} 
To obtain the proper large $N$ limit, the vector field is written as,
$$
\vec{\Phi}(\vec x, t) = (\sigma(\vec x,t), \vec{\pi}(\vec x,t)),
$$ 
with $\vec{\pi}$ an $N-1$-plet, and we write,
\begin{equation}
\sigma(\vec x,t) = \sqrt{N}\phi(t) + \chi(\vec x,t) \; \; ; \; \; \langle
\sigma(\vec x, t) \rangle= \sqrt{N}\phi(t) \; \; ; \; \; \langle \chi(\vec x,
t) \rangle = 0.
\label{largenzeromode} 
\end{equation}
To implement the large $N$ limit in a consistent manner, one may introduce an
auxiliary field as in\cite{largen,largenfrw}.  However, the leading order
contribution can be obtained equivalently by invoking the factorization,

\begin{eqnarray}
\chi^4 & \rightarrow & 6 \langle \chi^2 \rangle \chi^2 +\text{constant},
\label{larg1} \\ \chi^3 & \rightarrow & 3 \langle \chi^2 \rangle \chi,
\label{larg2} \\ \left( \vec{\pi} \cdot \vec{\pi} \right)^2 & \rightarrow &
2 \langle \vec{\pi}^2 \rangle \vec{\pi}^2 - \langle \vec{\pi}^2 \rangle^2+
{\cal{O}}(1/N), \label{larg3} \\ \vec{\pi}^2 \chi^2 & \rightarrow & \langle
\vec{\pi}^2 \rangle \chi^2 +\vec{\pi}^2 \langle \chi^2 \rangle,
\label{larg4} \\ \vec{\pi}^2 \chi & \rightarrow & \langle \vec{\pi}^2
\rangle \chi.  \label{larg5}
\end{eqnarray}

To obtain a large $N$ limit, we define,
\begin{equation} 
\vec{\pi}(\vec x, t) = \psi(\vec x, t)
\overbrace{\left(1,1,\cdots,1\right)}^{N-1}, \label{filargeN}
\end{equation} 
where the large $N$ limit is implemented by the requirement that,
\begin{equation}
\langle \psi^2 \rangle \approx {\cal{O}} (1) \; , \; \langle \chi^2 \rangle
\approx {\cal{O}} (1) \; , \; \phi \approx {\cal{O}} (1).
\label{order1}
\end{equation}
The leading contribution is obtained by neglecting the $ {\cal{O}} ({1}\slash
{N})$ terms in the formal limit. The resulting Lagrangian density is quadratic,
with a linear term in $\chi$.  The equations of motion in this case become,

\begin{equation}
\ddot{\phi}(t)+3H\dot{\phi}(t)+M^2\phi(t)+\frac{\lambda}{2}\phi^3(t)+
\frac{\lambda}{2}
\phi(t)\langle \psi^2(t)\rangle=0, \label{largezeromodeeqn}
\end{equation}
\begin{equation}
\langle \psi^2(t) \rangle = \int
\frac{d^3k}{(2\pi)^3}\frac{|f_k(t)|^2}{2}\coth\left[\frac{W_k}{2T_i}\right],
\label{largenfluc}
\end{equation}
with the mode functions, 
\begin{equation}
\left[\frac{d^2}{dt^2}+3H\frac{d}{dt}+\omega^2_k(t) \right]f_k(t)= 0,
\label{largenmodes}
\end{equation}
and the effective frequencies,
\begin{equation}
\omega^2_k(t) =\frac{k^2}{a^2(t)}+M^2(t) \; ,
\label{largenfreq}
\end{equation}
where,
\begin{equation}
M^2(t) =  M^2+ \frac{\lambda}{2}\phi^2(t)+
\frac{\lambda}{2}\langle \psi^2(t) \rangle \; .
\label{Ngranmass}
\end{equation}

The initial conditions are chosen to reflect the same physical situation as in
the Hartree case, that is, the instantaneous particle states of the Hamiltonian
at $t=0$ are in local thermodynamic equilibrium at some initial temperature
higher than the critical value.  Thus, as in the Hartree case but with
modified frequencies, the initial conditions at $t=0$ are chosen to describe
the instantaneous positive energy states,
\begin{equation}
f_k(0)= \frac{1}{\sqrt{W_k}} \; ; \; \dot{f}_k(0) = -i \sqrt{W_k} \; \;
; \; \; 
W_k= \sqrt{k^2+M^2_0}. \label{largenbc}
\end{equation}
We have maintained the same names for the mode functions and $M_0$ to avoid
cluttering of notation; their meaning for each case should be clear from the
context. Notice that the difference between the Hartree and large $N$ case is
rather minor.  The most significant difference is that, in the equations for
the zero modes, the Hartree case displays a factor 3 difference between the
tree level non-linear term and the contribution from the fluctuation as
compared to the corresponding terms in the large $N$ case.  The equations for
the mode functions are the same upon a trivial rescaling of the coupling
constant by a factor 3.

The particular case with $\phi (t) =0 $ is of interest, since by symmetry, it
is a fixed point of the dynamics of the zero mode and corresponds to the case
in which the phase transition occurs from the symmetric phase into the broken
phase in the absence of symmetry breaking perturbations or initial bias in the
field. In this important case, the Hartree and large $N$ mode equations are
identical after the coupling constant is rescaled by a factor of $3$, and our
conclusions will be {\em universal} in the sense that both the large $N$
approximation and the Hartree approximation describe the same non-perturbative
dynamics.  In the large $N$ limit we find the energy density to be given by,
\begin{eqnarray}
\frac{\varepsilon}{N} = & & \frac{1}{2}\dot{\phi}^2(t)+
\frac{\lambda}{8}\left(\phi^2(t)+\frac{2M^2}{\lambda}\right)^2 + \nonumber \\
&& \frac{1}{2} \int \frac{d^3k}{2(2\pi)^3 }\coth\left[\frac{W_k}{2T_i}\right]
\left[ |\dot{f}_k(t)|^2+ \omega^2_k(t)|f_k(t)|^2\right] -
\frac{\lambda}{8}\langle \psi^2(t) \rangle^2,
\label{largenenergy} 
\end{eqnarray}
and $(p+\varepsilon)/N$ has the same form as for the Hartree case above
(eq. (\ref{pplusehart})), but in terms of the mode functions obeying the large
$N$ equations. Again, it is straightforward to show that the bare energy is
covariantly conserved by using the equations of motion for the zero mode and
the mode functions.  Note that equation (\ref{hartcons}) also holds in the
large $ N $ limit.

A relation analogous to (\ref{hartcons}) holds for the integrands of $
\varepsilon $ and $ \varepsilon + p $ in eqs.(\ref{hartenergy}) and
(\ref{pplusehart}). Namely,
\begin{equation}
{\dot \varepsilon}_k(t) + 3 H \sigma_k(t) = \frac{d}{dt}[M^2(t)]\; |f_k(t)|^2
\; .
\label{consD}
\end{equation}
Here,
\begin{eqnarray}
\varepsilon_k(t) &=& |\dot{f}_k(t)|^2 + \omega^2_k(t)\; |f_k(t)|^2 \; , \cr \cr
\sigma_k(t) &=& 2\left[ |\dot{f}_k(t)|^2+ \frac{k^2}{3a^2(t)}|f_k(t)|^2\right]
\; .\label{epsdek}
\end{eqnarray}
Equation (\ref{consD}) will be used below to prove that the renormalized energy
is conserved.

\section{\bf Renormalized Equations of Motion}

Renormalization is a very subtle but important issue in gravitational
backgrounds\cite{vilford,birriel,anderson}. The fluctuation contribution
$\langle \psi^2(\vec x,t) \rangle$, the energy and the pressure all need to be
renormalized. The renormalization aspects in curved space times have been
discussed at length in the literature\cite{vilford,birriel,anderson} and has
been extended to the Hartree and large $N$ self-consistent approximations for
the non-equilibrium backreaction problem in\cite{frw,largenfrw}.

In terms of the effective mass term for the large $ N $ limit given by
(\ref{Ngranmass}) and the Hartree case, eq. (\ref{hartmass}), 
and defining the quantity,
\begin{equation}
B(t) \equiv a^2(t)\left(M^2(t)-{\cal{R}}/6 \right)\label{boft},
\end{equation} 
we find the following large $k$ behavior for the case of an {\em arbitrary}
scale factor $a(t)$ (with $a(0)=1$),
\begin{eqnarray}
|f_k(t)|^2 &=& \frac{1}{ka^2(t)}+ \frac{1}{2k^3
a^2(t)}\left[H^2(0)-B(t) \right] \cr \cr &+&
{1 \over {8 a(t)^2 \; k^5 }}\left\{ B(t)[ 3 B(t) - 2 H^2(0) ] + a(t)
\frac{d}{dt} \left[ a(t) {\dot B}(t) \right] + D_0 \right\} + {\cal{O}}(1/k^7)
\cr \cr |\dot{f}_k(t)|^2 &=&
\frac{k}{a^4(t)}+\frac{1}{ka^2(t)}\left[H^2(t)+\frac{H^2(0)}{2a^2(t)}+
\frac{1}{2}\left(M^2(t)-{\cal{R}}/6 \right) \right] \cr \cr & + & {1 \over {8
a(t)^4 \; k^3 }}\left\{ - B(t)^2 - a(t)^2 {\ddot B}(t) + 3 a(t){\dot a}(t)
{\dot B}(t) - 4 {\dot a}^2(t) B(t) \right. \cr \cr &+& \left.  2 H^2(0) [ 2
{\dot a}^2(t) + B(t) ] + D_0 \right\} +{\cal{O}}(1/k^5). \label{largekf}
\end{eqnarray}
The constant $ D_0 $ depends on the initial conditions and is unimportant
for our analysis.

Although the divergences can be dealt with by dimensional regularization, this
procedure is not well suited to numerical analysis.  We will make our
subtractions using an ultraviolet cutoff constant in {\em physical
coordinates}. This guarantees that the counterterms will be time
independent. The renormalization then proceeds much in the same manner as in
reference\cite{frw}; the quadratic divergences renormalize the mass and the
logarithmic terms renormalize coupling constant and the coupling to the Ricci
scalar.  The logarithmic subtractions can be neglected because of the coupling
$\lambda \approx 10^{-12}$.  Using the Planck scale as the cutoff and the
inflaton mass $m_R$ as a renormalization point, these terms are of order
$\lambda \ln[M_{pl}/m_R] \leq 10^{-10}$, for $m \geq 10^9 \mbox{GeV}$. An
equivalent statement is that for these values of the coupling and inflaton
masses, the Landau pole is beyond the physical cutoff $M_{pl}$.  Our relative
error in the numerical analysis is of order $10^{-8}$, therefore our numerical
study is insensitive to the logarithmic corrections. Though these corrections
are fundamentally important, numerically they can be neglected. Therefore, in
what follows, we will neglect logarithmic renormalization and subtract only
quartic and quadratic divergences in the energy and pressure, and quadratic
divergences in the fluctuation contribution.

Using the large $k$ behavior of the mode functions, we find the finite
temperature factor to be\cite{frw},
\begin{equation}
\int {d^3k}\frac{|f_k(t)|^2}{\exp[\frac{W_k}{T_i}]-1} = 
\frac{1}{24}\left[\frac{T^2_i}{a^2(t)}\right]\left[1+{\cal{O}}(M_0/T_i)+ 
{\cal{O}}\left(\ln(M_0/T_i)\right)+\cdots \right], \label{hiT}
\end{equation}
where we have used $T_i > T_c >> M_0$. The first term in (\ref{hiT}) arises
from the $ 1/[ka^2(t)] $ in the asymptotic limit (\ref{largekf}). For the large $N$
case, we find the renormalization condition,
\begin{equation}
-m^2_B+\frac{\lambda_B}{2}\phi^2(t)+\xi_B{\cal{R}}+\frac{\lambda_B}{2}
 \langle \psi^2(t) \rangle_B = 
m^2_R(t,T_i)+\frac{\lambda_R}{2}\phi^2(t)+\xi_R{\cal{R}}+\frac{\lambda_R}{2}
 \langle \psi^2(t) \rangle_R, \label{renor}
\end{equation}
with the effective time dependent mass term in de Sitter space time, and
subtracted fluctuation contribution\cite{frw},
\begin{eqnarray}
m^2_R(t,T_i) &=& m^2_R\left[\frac{T^2_i}{T^2_c}e^{-2Ht}-1\right] \; , \;
\langle \psi^2(t) \rangle_R = \frac{1}{4\pi^2}\int_0^{\infty} k^2 \, dk 
\coth\left[\frac{W_k}{2T_i}\right]
 \left\{ |f_k(t)|^2 \right. \cr \cr &-&
\left. \frac{1}{ka^2(t)} - \frac{\theta(k-\kappa)}{2k^3
a^2(t)}\left[H^2(0)-a^2(t)\left(M^2(t)-{\cal{R}}/6 \right) \right] \right\},
\end{eqnarray}
where $\kappa$ is a renormalization point, which for convenience will be chosen
as $\kappa= |m_R|$, i.e. the renormalized inflaton mass.  The Hartree case is
similar but with a rescaling of the coupling constant by a factor 3.  We will
fix the renormalized coupling to the Ricci scalar to be $\xi_R=0$, thus
ensuring minimal coupling in the renormalized theory.

Assuming that $T_i/T_c \approx {\cal O}(1)$, we see that the phase transition
occurs within the first few e-folds of inflation.  When the temperature falls
below the critical value, the effective mass becomes negative.  As will be seen
explicitly below, when this occurs, long-wavelength modes become unstable and
grow. Local thermodynamic equilibrium will set in again if the contribution
from the quantum fluctuations can grow and adjust to compensate for the
negative mass terms on the same time scales as that in which the temperature
drops. However, as discussed below, for very weak coupling the important time
scales for the non-equilibrium fluctuations are of the order of $ [H/m^2_R]
\ln[1/\lambda] $, which are much longer than the time it takes for the
temperature to drop well below the critical value to practically zero. Thus,
the non-equilibrium dynamics will proceed as if the phase transition occured
via a ``quench'', that is with an effective mass term,
\begin{equation}
m^2_{eff}(t)= m^2_i\theta(t_i-t)-m^2_R\theta(t-t_i)
\quad ; \quad m^2_i=  m^2_R\left[\frac{T^2_i}{T^2_c}-1\right] >0 \, . 
 \label{quench}
\end{equation}
Therefore we choose the initial conditions on the mode functions at $t_i=0$ to
be given in terms of the effective mass,
\begin{equation}
M^2_0= m^2_R\left[ \frac{T^2_i}{T^2_c}-1\right]+\frac{\lambda_R}{2}\phi^2(0)
\quad ; \quad \frac{T_i}{T_c} >1,
\label{minitial}
\end{equation}
with $\lambda_R \rightarrow 3 \lambda_R$ for the Hartree case.

We introduce the following dimensionless quantities and definitions,
\begin{equation}
\tau = m_R t \quad ; \quad h= \frac{H}{m_R} \quad ; \quad q=\frac{k}{m_R} \; ,
\label{dimvars1}
\end{equation}
\begin{equation}
r = \frac{T_i}{T_c} \quad ; \quad {\cal{T}}= \frac{T_i}{m_R} \quad; \quad
\omega_q = \frac{W_k}{m_R} \quad ; \quad g= \frac{\lambda}{8\pi^2} \; ,
\label{dimvars2}
\end{equation}
\begin{equation}
\eta^2(\tau) = \frac{\lambda}{2m^2_R} \; \phi^2(t)
\quad ; \quad  g\Sigma(\tau) = \frac{\lambda}{2m^2_R}\; \langle \psi^2(t)
\rangle_R \; ,
\label{dimvars3}
\end{equation}

\begin{equation}
f_q(\tau) \equiv \sqrt{m_R} \; f_k(t) \quad ; \quad K = {{\kappa}\over {m_R}}
\; .
\label{dimensionlessdefs}
\end{equation}

With $\xi_R=0$  and $ K = 1$ the equations of motion become:

\vspace{2mm} 

{\bf Large N:}
\begin{equation}
\ddot{\eta}+ 3h \; \dot{\eta}-\eta+\eta^3+ g\Sigma(\tau)\eta = 0
\label{dimlesslargenzero}
\end{equation}

\begin{equation} 
\left[\frac{d^2}{d \tau^2}+3h \frac{d}{d
\tau}+\frac{q^2}{a^2(\tau)}-1+\eta^2+g\Sigma(\tau) \right]f_q(\tau)=0
\label{dimlesslargenmodes}
\end{equation}
\begin{equation}
 f_q(0)=\frac{1}{\sqrt{\omega_q}} \quad ; \quad
\dot{f}_q(0) = -i \sqrt{\omega_q} \quad ; \quad \omega_q=
\sqrt{q^2+r^2-1+\eta^2(0)}
\label{largenincon}
\end{equation}

{\bf Hartree:}

\begin{equation}
\ddot{\eta}+ 3h \; \dot{\eta}-\eta+\eta^3+ 3g\Sigma(\tau)\eta = 0
\label{dimlesshartzero}
\end{equation}

\begin{equation} 
\left[\frac{d^2}{d\tau^2}+3h
\frac{d}{d\tau}+\frac{q^2}{a^2(\tau)}-1+3\eta^2+3g\Sigma(\tau)
\right]f_q(\tau)=0 \label{dimlesshartmodes}
\end{equation}
\begin{equation}
 f_q(0)=\frac{1}{\sqrt{\omega_q}} \quad ; \quad \dot{f}_q(0) = -i
\sqrt{\omega_q} \quad ; \quad \omega_q= \sqrt{q^2+r^2-1+3\eta^2(0)}
\label{hartinconds}
\end{equation}
The dots stand for derivatives with respect to $\tau$, and in both cases,
\begin{equation}
g\Sigma(\tau) = \int_0^{\infty} q^2 \, dq \, 
\coth\left[\frac{\omega_q}{2 {\cal T}}\right]
\left\{ |f_q(\tau)|^2 -
\frac{\theta(q-1)}{2q^3 a^2(t)}\left[h^2(0)-{{a^2(t)}\over
{m_R^2}}\left(M^2(t)-{\cal{R}}/6 \right) \right] \right\}.
\label{gsigdimless}
\end{equation}

Numerically, the most significant contribution to $\langle \psi^2 \rangle$
arises from low wavevectors $q \leq 10-20$ in all of the cases studied (see
figures 1.e and 1.f below). Since $T_i > T_c = \sqrt{24 m^2_R/ \lambda}$, $q <<
{\cal T}$ and $M_0<<T_i$ for these low momentum modes, they are ``classical''
and we can approximate,
\begin{equation}
\coth\left[\frac{\omega_q}{2 {\cal T}}\right] \approx \frac{2 {\cal
T}}{\omega_q} = \frac{2r}{\omega_q} \sqrt{\frac{24}{\lambda}}. \label{hiTlim}
\end{equation}

In this approximation, the fluctuation contribution becomes,
\begin{eqnarray}
g\Sigma(\tau) &=& \bar{g}\int \frac{q^2 dq}{\omega_q}\left\{|f_q(\tau)|^2-
\frac{\theta(q-1)}{2q^3 a^2(t)}\left[h^2(0) - {{a^2(t)}\over
{m_R^2}}\left(M^2(t)-{\cal{R}}/6 \right) \right] \right\} \; , \cr \cr \bar{g}
&=& 2rg \sqrt{\frac{24}{\lambda}} = 1.103 \, r \sqrt{g}= 0.124 \, r
\sqrt{\lambda} \; . \label{gsigma}
\end{eqnarray} 

The renormalized energy density and pressure for the large $N$ case are
given by,

\begin{eqnarray}
\frac{\varepsilon}{N} &=& \frac{2m^4_R}{\lambda}\left\{
\frac{\dot{\eta}^2}{2}+\frac{1}{4}(\eta^2-1)^2+\frac{\bar{g}}{2} \int \frac{q^2
dq}{\omega_q}\left[|\dot{f}_q(\tau)|^2 + \omega^2_q(\tau)|f_q(\tau)|^2- {{2 \,
q}\over {a(t)^4 }} - {{\alpha(t)} \over q} \right. \right. \cr \cr &-&
\left. \left.  {{\theta(q-1)\; \beta(t)}\over{q^3}} \right]
-\frac{g}{4}\Sigma(\tau) \right\}, \label{finener}
\end{eqnarray}
\begin{equation}
\frac{p+\varepsilon}{N} = \frac{2m^4_R}{\lambda}\left\{ \dot{\eta}^2+
\bar{g}\int \frac{q^2 dq}{\omega_q} \left[|\dot{f}_q(\tau)|^2 +
\frac{q^2}{3a^2(\tau)}|f_q(\tau)|^2 - {{4q}\over {3 a(t)^4 }} - {{\gamma(t)}
\over q} - \frac{\theta(q-1)}{q^3}\, \delta(t) \right] \right\},
\label{presfin}
\end{equation}
with,
\begin{equation}
\omega^2_q(\tau) = \frac{q^2}{a^2(\tau)}-1+\eta^2(\tau)+g\Sigma(\tau).
\label{frequ}
\end{equation} 
and $B(t)$ is given by eq.(\ref{boft}). 
The coefficients $ \alpha(t), \; \beta(t), \; \gamma(t) $ and $ \delta(t) $ are
obtained from the asymptotic behaviour of $\varepsilon_k(t)$ and $\sigma_k(t)$
(\ref{epsdek}-\ref{largekf}). We find,
\begin{eqnarray}
\alpha(t) &=& { 1 \over {m_R^2\, a(t)^4 }}\left[ {\dot a}^2(t) + a^2(t) \,
M(t)^2 + H(0)^2 \right], \cr \cr \beta(t) &=& { 1 \over {4 \, m_R^4\, a(t)^4
}}\left\{ B^2(t) \,+ 2 a(t) {\dot a}(t) {\dot B}(t) \right. \cr \cr & - &
\left. 2 [ {\dot a}^2(t) + a^2(t) \, M(t)^2 ][B(t) - H(0)^2 ] + D_0\right\},
\cr \cr \gamma(t) &=& { 1 \over {3 \, m_R^2\, a(t)^4 }}\left[ B(t) + 3 {\dot
a}^2(t) + 2 H(0)^2\right] , \\ \cr \delta(t) &=& -{ 1 \over {12\, m_R^4\,a(t)^4
}}\left\{ a^2(t) \, {\ddot B}(t)-5 \, a(t) \, {\dot a}(t) \, {\dot B}(t)
\right. \cr \cr &+& \left. 6 \, {\dot a}^2(t) \, [ B(t)- H(0)^2] - 2 \, H(0)^2
B(t)-2D_0\right\}.  \nonumber
\end{eqnarray}

For the Hartree case (setting $N=1$) the only changes are that we take $g\Sigma
/4 \rightarrow 3g\Sigma /4 $ in the last term in the energy, and use the
frequencies,
\begin{equation}
\omega^2_q(\tau) = \frac{q^2}{a^2(\tau)}-1+3\eta^2(\tau)+3g\Sigma(\tau),
\label{harfrequ}
\end{equation} 
and that the initial conditions on the mode functions are given by
(\ref{hartinconds}).

In order to prove that the renormalized energy and pressure fullfils the
continuity equation (\ref{hartcons}), we need to study the properties of the
subtracted terms in equations (\ref{finener})-(\ref{presfin}). Inserting the
asymptotic behaviour of $\varepsilon_k(t), \; \sigma_k(t) $ and $|f_k(\tau)|^2$
for large $ k $ in (\ref{consD}), we find,
\begin{eqnarray}
{\dot  \alpha}(t) + 6 H \;  \gamma(t) = {1 \over {a^2(t)}}\;
\frac{d}{dt}[M^2(t)]\; ,  \cr \cr
{\dot \beta}(t) + 6 H \; \delta(t) = {1 \over {2 a^2(t)}}\; [H(0)^2
- B(t) ]\frac{d}{dt}[M^2(t)] \; .
\end{eqnarray}
Using these relations and (\ref{consD}), it is 
straightforward to show that the renormalized energy and pressure
given by eqs. (\ref{finener})-(\ref{presfin}) satisfy the continuity
equation,
$$
\dot{\varepsilon}+3H(p+\varepsilon)=0 \; .
$$
A noteworthy point is that when the cutoff is kept fixed in {\em physical}
coordinates, upon taking the time derivative in the integrals, there
is a contribution from the upper limit of the integrals. However the
subtractions guarantee that in the formal limit when the cutoff is taken
to infinity this contribution vanishes.  While the existence of the Landau
pole beyond the Planck mass restricts taking the cutoff to infinity in
this effective theory, we find that any contributions from the 
upper limit are numerically
small.  In our numerical evolution, the energy density is covariantly 
conserved to one part in $10^7$.

From the evolution of the mode functions that determine the quantum
fluctuations, we can study the growth of correlated domains with the equal time
correlation function,

\begin{eqnarray}\label{correlator}
S(\vec{x},t) & = & \langle\psi(\vec{x},t)\psi(\vec{0},t)\rangle, \\ \nonumber &
= & \int\frac{d^3k}{(2\pi)^3}e^{i\vec{k}\cdot\vec{x}}
\frac{|f_k(t)|^2}{2}\coth\left(\frac{W_k }{2T_i}\right),
\end{eqnarray}
which can be written in terms of the  power spectrum of quantum fluctuations,
\begin{equation}
{\cal{S}}(k,t) = \frac{{\cal S}(q,\tau)}{m_R} \quad ; \quad {\cal S}(q,\tau) =
{|f_q(\tau)|^2}\coth\left(\frac{\omega_q }{2{\cal T}}\right).
\label{powerspectrum}
\end{equation}
It is convenient to define the dimensionless correlation function,
\begin{equation}
{\cal S}(\rho,\tau) = \frac{S(|\vec x|, t)}{m^2_R}= \frac{1}{{4\pi^2 }\rho}
\int qdq \sin[q\rho] {\cal{S}}(q,\tau) \; ; \rho= m_R|\vec x|.  \label{rhocorr}
\end{equation}

We now have all the ingredients to study the particular cases of interest.

\section{Evolution for $\phi(0)= \dot{\phi}(0)=0$}

\subsection{Analytical Results}

We begin by considering the situation in which the expectation value of the
inflaton field sits atop the potential hill with zero initial velocity. This
situation is expected to arise if the system is initially in local
thermodynamic equilibrium  an initial
temperature larger than the critical temperature and cools down
through the critical temperature in the absence of an external field or bias.

The order parameter and its time derivative vanish in the
local equilibrium high
temperature phase, and this condition is a fixed point of the evolution
equation for the zero mode of the inflaton.  There is no rolling of the
inflaton zero mode in this case, although the fluctuations will grow and will be responsible for the dynamics.

We can understand the early stages of the dynamics analytically as follows. For
very weak coupling and early time we can neglect the backreation in the mode
equations, which in both the large $N$ and Hartree cases become,

\begin{equation} 
\left[\frac{d^2}{d\tau^2}+3h
\frac{d}{d\tau}+\frac{q^2}{a^2(\tau)}-1\right]f_q(\tau)=0, \label{earlytime}
\end{equation}
\begin{equation}
f_q(0)=\frac{1}{\sqrt{\omega_q}} \; ; \quad \dot{f}_q(\tau) = -i
\sqrt{\omega_q} \; ; \quad \omega_q= \sqrt{q^2+r^2-1} \; .
\end{equation}
The solutions are of the form,
\begin{equation}
f_q(\tau) = \exp[-\frac{3}{2}h\tau] \left\{a(q)\; J_{\nu}(z)+b(q)\;
J_{-\nu}(z) \right\} \; ; \;
z=\frac{q}{h}\exp[-h\tau] 
\; ; \; \nu = \sqrt{\frac{1}{h^2}+\frac{9}{4}}, \label{bessel}
\end{equation}
where the coefficients $a(q)$ and $b(q)$ are determined by the initial
conditions:
\begin{equation}
b(q) = - {{\pi \, q}\over { 2 h \, \sin{\nu \pi} }} \; \left[ {{i \omega_q -
\frac32 \, h }\over q} \; J_{\nu}\left(\frac{q}{h}\right) - 
J'_{\nu}\left(\frac{q}{h}\right) \right],
\label{coefb}
\end{equation}
\begin{equation}
a(q) =   {{\pi \, q}\over { 2 h \, \sin{\nu \pi} }} \; \left[ {{i \omega_q -
\frac32 \, h }\over q} \;  J_{-\nu}\left(\frac{q}{h}\right) -  
J'_{-\nu}\left(\frac{q}{h}\right)\right] \;. \label{coefa}
\end{equation}

For long times, $e^{h\tau}\geq q/h$, these mode functions grow exponentially,
\begin{equation}\label{Uasi}
f_q(\tau) \simeq b(q) \; J_{-\nu}(z) \simeq {{b(q)} \over {\Gamma(1-\nu)}}
\; \left( {{2h\, }\over q}\right)^{\nu}e^{(\nu-3/2)h \tau}   \;.
\end{equation}

The Bessel functions appearing in the expression for the modes $ f_q(\tau) $
can be approximated by their series expansion,
\begin{equation}
f_q(\tau) = \frac12 \left[ 1 + \frac{1}{\nu} \; \left( \frac32 -{{q^2}\over{4
h^2}} - i {{\omega_q}\over h} \right) + {\cal O}\left(\frac1{\nu^2}\right) \right] \; 
e^{(\nu - 3/2) h\tau} \; .
\end{equation}
This is an expansion in powers of $ q^2/(\nu h^2) $ and we conclude that $
g\Sigma(\tau) $ is dominated by the modes with $q \leq \sqrt{h}$.

The integral for $g\Sigma(\tau)$ can be approximated by keeping only the modes
$ q \leq f \sqrt{h} $, where $ f $ is a number of order one, and by neglecting
the subtraction term which will cancel the contributions from high
momenta. Numerically, even with the backreaction taken into account, the
integral is dominated by modes $q \leq f \approx 10-20$ in all of the cases
that we studied (see fig. (1.e) below).

The contribution to the fluctuations from these unstable modes is:
\begin{equation}\label{estim}
 g\Sigma(\tau) \simeq \sqrt{{\lambda}\over 6}\, {{f^3\,h^{3/2} \, r m^2_R}
\over { 4 \pi^2 \, \, M_0^2 }}\; \left(1 + {{ M_0^2}\over { m^2_R}}\right) \;
e^{(2\nu-3)h\tau} \; ,
\end{equation}
where again, we have taken the high temperature limit, $ T_i \sim T_c \gg m_R
$.

From this equation, we can estimate the value of $\tau_s$, the ``spinodal
time'', at which the contribution of the quantum fluctuations becomes
comparable to the contribution from the tree level terms in the equations of
motion. This time scale is obtained
 from the condition $g\Sigma(\tau_s) = {\cal O}(1)$:
\begin{equation}
\tau_s \simeq -\frac{1}{(2\nu-3)h}\ln\left[\sqrt{{\lambda}\over 6}
{{f^3\,h^{3/2}}\over { 4 \pi^2 \, m_R M_0^2 }}\;\frac{T_i}{T_c} 
\left(1 + {{ M_0^2}\over { m_R^2}}\right)\right]\label{spinoest} ,
\end{equation}
which is in good agreement with our numerical results, as will become clear
below (see figures (1.a), (2.a) and (3.b)). For values of $h \geq 1$,
which, as argued
below, lead to the most interesting case, an estimate for the spinodal time is,
\begin{equation} \label{tslargeh}
\tau_s \simeq  \frac{3h}{2} \ln[1 / \sqrt{\lambda}] + {\cal{O}}(1)
\label{spinotime}
\end{equation}
which is consistent with our numerical results (see figure (1.a)).

For $\tau > \tau_s$, the effects of backreaction become very important, and the
contribution from the quantum fluctuations competes with the tree level terms
in the equations of motion, shutting-off the instabilities. Beyond $\tau_s$,
only a full numerical analysis will capture the correct dynamics.

It is worth mentioning that had we chosen zero temperature initial conditions,
then the coupling $\bar{g} \rightarrow g $ (see (\ref{gsigma})) 
and the estimate for the spinodal
time would have been,
\begin{equation}
\tau_s \simeq \frac{3h}{2} \ln[1/{\lambda}] + {\cal{O}}(1), \label{spinotimeT0}
\end{equation}
that is, roughly a factor 2 larger than the estimate for which the de Sitter
stage began at a temperature above the critical value. Therefore
(\ref{spinotime}) represents an {\em underestimate} of the spinodal time scale
at which fluctuations become comparable to tree level contributions.

The number of e-folds ocurring during the stage of growth of spinodal
fluctuations is therefore,
\begin{equation}
{\cal{N}}_e \approx \frac{3h^2}{2} \ln[1/ \sqrt{\lambda}], \label{efolds}
\end{equation}
or in the zero temperature case,
\begin{equation}
{\cal{N}}_e \approx \frac{3h^2}{2} \ln[1/ \lambda], \label{efoldsT0}
\end{equation}
which is a factor 2 larger.  Thus, it becomes clear that with $\lambda \approx
10^{-12}$ and $h \geq 2$, a required number of e-folds, ${\cal{N}}_e \approx
100$ can easily be accomodated before the fluctuations become large,
modifying the dynamics and the equation of state.

The implications of these estimates are important.  The first conclusion drawn
from these estimates is that a ``quench'' approximation is well justified (see
figure (1.a)). While the temperature drops from an initial value of a few
times the critical temperature to below critical in just a few e-folds, the
contribution of the quantum fluctuations needs a large number of e-folds
to grow
to compensate for the tree-level terms and overcome the instabilities. Only for
a strongly coupled theory is the time scale for the quantum fluctuations to
grow short enough to restore local thermodynamic
equilibrium during the transition.

The second conclusion is that most of the growth of spinodal fluctuations
occurs during the inflationary stage, and with $\lambda \approx 10^{-12}$ and
$H \geq m_R$, the quantum fluctuations become of the order of the tree-level
contributions to the equations of motion within the number of e-folds necessary
to solve the horizon and flatness problems.  
Since the fluctuations grow to become
of the order of the tree level contributions at times of the
order of this time scale, for larger times they will
modify the equation of state substantially and will be shown to provide a
graceful exit from the inflationary phase within an acceptable number of
e-folds.

For $\tau < \tau_s$, when the contribution from the renormalized quantum
fluctuations can be ignored, the Hubble constant is given by the classical
contribution to the energy density. In terms of the dimensionless quantities
introduced above (\ref{dimvars3}), we have,
\begin{equation}
H= \frac{16\pi m^4_R}{3\lambda
M^2_{pl}}\left[\frac{\dot{\eta}^2}{2}+\frac{1}{4}(\eta^2-1)^2\right].
\label{hubbleconst}
\end{equation}
In the situration we consider here, with $\dot{\eta}=\eta=0$, the condition
that $h \geq 2$ for $\lambda \simeq 10^{-12}$ translates into $m_R \simeq
10^{13}\mbox{ GeV }$, which is an acceptable bound on the inflaton mass.

To understand more clearly whether or not the effect of quantum fluctuations
and growth of unstable modes during the inflationary phase transition can
provide a graceful exit scenario, we must study in detail the contribution to
the energy and the equation of state of these quantum fluctuations.

Although we are working in a fixed de Sitter background, the energy and
pressure will evolve dynamically. A measure of the backreaction effects of
quantum fluctuations on the dynamics of the scale factor is obtained from
defining the `effective Hubble constant',
\begin{equation}
{\cal{H}}^2(\tau) = \frac{8\pi}{3M^2_{pl}}\; \varepsilon(\tau). \label{hub}
\end{equation} 
Therefore, the quantities,
\begin{equation}
\frac{{\cal{H}}(\tau)}{{\cal{H}}(0)}= \sqrt{\frac{
\varepsilon(\tau)}{\varepsilon(0)}},
\label{hubratio}
\end{equation}
and
\begin{equation}
\frac{\dot{\cal{H}}(\tau)}{{\cal{H}}^2(\tau)} = -\frac{3}{2}
\left[1+\frac{p(\tau)}{\varepsilon(\tau)}\right], \label{hdot}
\end{equation}
give dynamical information of the effects of the backreaction of the quantum
fluctuations on the dynamics of the scale factor.  Whenever
$p(\tau)+\varepsilon(\tau) \neq 0$, ${\cal{H}}(\tau)/{\cal{H}}(0) \neq 1$, or
$\dot{\cal{H}}(\tau)/ {\cal{H}}^2(\tau) \neq 0$, the backreaction from the
quantum fluctuations will dramatically change the dynamics of the scale factor,
and it will no longer be consistent to treat the scale factor as fixed. When
${\cal H}(\tau)/ {\cal H}(0) \ll 1$ the de Sitter era will end.

From this point onwards only a full treatement of the backreaction, {\em
including} the correct dynamics of the scale factor, will describe the physics.
The time scale on which the quantum fluctuations will begin to influence the
dynamics of the scale factor is of the order of the spinodal time estimated
above, since the contribution of the quantum fluctuations becomes comparable
to the tree level terms and modifies the equation of state.

Therefore, there is the possibility that the growth of quantum fluctuations can
provide a graceful exit from the inflationary phase, even when the zero mode
{\em does not roll}. The parameters should be chosen in such a way so that the
requisite 60 or more e-folds of expansion take place before the spinodal time.
From the estimates provided above, this is relatively easy to accommodate with
reasonable values of the inflaton mass and for the weak coupling that is
usually assumed in inflationary models.

\subsection{\bf Numerical Analysis}

We now solve the large $N$ set of equations (\ref{dimlesslargenmodes})
numerically, with the initial conditions (\ref{largenincon}), taking
$\dot{\eta}(0) = \eta(0) =0$.

The numerical code is based on a fourth order Runge-Kutta algorithm for the
differential equation and an 11-points Newton-Cotes algorithm for the integral,
with a typical relative errors $10^{-9}$ in the differential equation and
in the integrals. We have tested for cutoff insensitivity with
cutoffs $q_{max} \approx 50 \; ; 100 \; ; 150$ with no appreciable variation in
the numerical results.  The reason for this cutoff insensitivity is due to the
fact that only long-wavelength modes grow in amplitude to become
non-perturbatively large, whereas the short-wavelength modes always have
perturbatively small amplitudes.  We have chosen $r=T_i/T_c =2$ as a
representative value and $\lambda= 10^{-12}$. The insensitivity on the value of
the cutoff confirms that the high-temperature limit (\ref{hiTlim}) is
warranted.

As argued previously, for $\lambda \approx 10^{-12}$, the cosmologically
interesting time scales for the spinodal instabilities to grow during say the
first 60-100 e-folds of inflation occur for $h \geq 1$, leading to $H \geq m_R
\geq 10^{13} \mbox{ GeV }$ which is a phenomenologically acceptable 
range for the Hubble constant during the inflationary stage.

Figure (1.a) shows the contribution from the quantum fluctuations,
$g\Sigma(\tau)$ vs. $\tau$ for $\lambda = 10^{-12} \; ; \; T_i/T_c=2 \; ; \;
h=2 \; ; \; \eta(0)=0\; ; \;\dot{\eta}(0)=0$. The quantum fluctuations, as
measured by $g\Sigma(\tau)$, grow to be of order 1 in a time scale $\tau
\approx 40$ which is the time scale predicted by the early time estimate
(\ref{spinotime}). Figure (1.b) shows $\left[p(\tau)+\varepsilon(\tau)\right]
\lambda/(2m^4_R)$ vs. $\tau$ for the same values of figure (1.a). 
Initially, $p=-\varepsilon$ and the quantity $p+\varepsilon$ is zero.
At the spinodal time, there is a change in the equation of state, causing
$p+\varepsilon$ to grow.  However for late times, the energy density and pressure
are each redshifted away such that the sum again approaches zero.  We have 
checked numerically that the energy is covariantly conserved, obeying the
relation $\dot{\varepsilon}+3H(p+\varepsilon)=0$ to our numerical accuracy of 
one part in $10^7$.
Figures (1.c,d) show ${\cal{H}}(\tau)/{\cal{H}}(0)$ and
$\dot{{\cal{H}}}(\tau)/{\cal{H}}^2(\tau)$ vs. $\tau$ respectively. These
figures show clearly that when the spinodal quantum fluctuations become
comparable to the tree level contribution to the equations of motion, the
backreaction on the scale factor becomes fairly large. At this point, the
approximation of keeping a fixed background breaks down and the full
self-consistent dynamics will have to be studied. At this time, the
inflationary stage basically ends since ${\cal{H}}$ is no longer constant. This
occurs for $\tau \approx 40$ giving about 80 e-folds of inflation during the
time in which ${\cal{H}}$ is approximately constant and equal to $H$. Therefore,
this new mechanism of spinodal fluctuations, with the zero mode sitting atop
the potential hill provides a graceful exit of the inflationary era without any
further assumptions on the evolution of the scalar field.

These fluctuations translate into an amplification of the power spectrum
at long
wavelengths for $q \approx h$. To see this clearly we plot $g
{\cal{S}}(q,\tau)$ with ${\cal{S}}(q,\tau)$ given by eq. (\ref{powerspectrum})
vs. $q$ for $\tau= 60$ in fig. (1.e). This quantity is very small, because of
the coupling constant in front, but for $\tau \approx \tau_s$ it grows to be of
order one for long wavelengths (see also fig. (1.h)) 
and vanishes very fast for $q > 10$. The
integral in $g\Sigma(\tau)$ is dominated by these long wavelengths that become
non-perturbatively large, whereas the contribution from the short wavelengths
remains always perturbatively small. This is the justification for the
approximations performed early that involved only the long-wavelength modes and
cutoffs of order $\sqrt{h}$. The equal time spatial
correlation function given by
eq. (\ref{rhocorr}) can now be computed explicitly. Figure (1.f) shows $S(\rho
; \tau)/S(0;\tau)$ as a function of $\rho$ for $\tau \ge 2$. We {\em define}
the correlation length $\xi(\tau)$ as the value of $\rho$ for which the ratio
is $1/e$. Figure (1.g) shows $\xi(\tau)$; notice that the correlation length
saturates to a value $\xi(\infty) \approx 1/h$, and that the correlated
regions are of horizon size.

We have performed numerical analysis varying $h$ with the same values of
$\lambda$ and for the same initial conditions, and found that the only
quantitative change is in the time scale for $g\Sigma(\tau)$ to be of order
one.  We find that the spinodal time scale grows almost linearly with $h$ and
its numerical value is accurately described by the estimate (\ref{spinotime}).
The case in which the Hubble constant is $h=0.1$ is shown explicitly.  Figure
(2.a) shows $g\Sigma(\tau)$, which demonstrates the oscillatory behavior 
similar to what is seen in Minkowski space\cite{boyrehe}.  The correlation
length, $\xi(\tau)$, is shown in fig. (2.b); its asymptotic value is again
approximately given by $1/h$.

\subsection{\bf The late time limit}

For times $\tau > \tau_s \approx 40$ (for the values of the parameters used in
figures (1)) we see from figures (1.a,h) that the dynamics freezes out. The
fluctuation, $g\Sigma(\tau) =1$, and the mode functions effectively describe
free, minimally coupled, massless particles.  The sum rule,
\begin{equation}
-1+g\Sigma(\infty) =0, \label{sumrule1}
\end{equation}
is obeyed exactly in the large $N$ limit as in the Minkowski
case\cite{boyrehe}.

For the Hartree case $g \rightarrow 3g$, but the physical phenomena are the
same, with the only difference that the sum rule now becomes $g\Sigma(\infty) =
1/3$. We now show that this value is a self-consistent solution of the
equations of motion for the mode functions, and the {\em only} stationary
solution for asymptotically long times.

In the late time limit, the effective time dependent mass term,
$-1+\eta^2+g\Sigma$, in the equation for the mode functions,
(\ref{dimlesslargenmodes}), vanishes (in this case with $\eta =0$).  Therefore,
the mode equations (\ref{dimlesslargenmodes}) asymptotically become,
\begin{equation} 
\left[\frac{d^2}{d \tau^2}+3h \frac{d}{d \tau}+\frac{q^2}{a^2(\tau)}
\right]f_q(\tau)=0 \; .  \label{dimlesslargenmod}
\end{equation}
The general solutions are given by,
\begin{equation}\label{asi}
f_q^{asy}(\tau)= \exp\left[-\frac{3}{2}h \tau \right] \left[d(q) \;
J_{3/2}\left( \frac{q}{h}e^{-h\tau}\right) + c(q) \; N_{3/2}\left(
\frac{q}{h}e^{-h\tau}\right)\right] \; ,
\end{equation}
where $ J_{3/2}(z) $ and $ N_{3/2}(z) $ are the Bessel and Neumann functions,
respectively.  The coefficients, $ d(q) $ and $ c(q) $, can be computed for
large $ q $ by matching $ f_q^{asy}(\tau) $ with the WKB approximation to the
exact mode functions $ f_q(\tau) $ that obey the initial conditions
(\ref{largenincon}). The WKB approximation to $f_q(\tau) $ has been computed
in ref.\cite{frw}, and we find for large $q $,
\begin{equation}\label{abdek}
d(q) = \sqrt{{\pi\, q}\over {2\, h}}\; \left[ 1 - {i\over q}(h + \Delta) +
{\cal O}(q^{-2}) \right]\;e^{-iq/h} + \sqrt{{\pi\, h}\over {8\, q}}\; \left[ 1
+ {\cal O}\left({1\over q}\right) \right]\;e^{iq/h} \; ,
\end{equation}
\begin{equation}
c(q) = -i\sqrt{{\pi\, q}\over {2\, h}}\; \left[ 1 - {i\over q}( h + \Delta) +
{\cal O}(q^{-2}) \right]\;e^{-iq/h} + i\sqrt{{\pi\, h}\over {8\, q}}\; \left[
1 + {\cal O}\left({1\over q}\right) \right]\;e^{iq/h} \; ,
\end{equation}
where
\begin{equation}
\Delta \equiv \int_0^{\infty} d\tau \; e^{h\tau} \; M^2(\tau) \; .
\end{equation}

In the $ \tau \to \infty $ limit, we have for fixed $ q $,
\begin{equation}\label{uasi}
f_q^{asy}(\tau) \buildrel{ \tau \to \infty}\over=  - \sqrt{2 \over
{\pi}}\left({h \over q }\right)^{3/2}\; c(q) \; \; .
\end{equation}
which are independent of time asymptotically, and explains why the power
spectrum of quantum fluctuations freezes at times larger than the
spinodal. This behavior is confirmed numerically: fig. (1.h) shows
$\ln\left[|f_q(\tau)|^2\right]$ vs. $\tau$ for $q=0,4,10$. Clearly at early
times the mode functions grow exponentially, and at times of the order of
$\tau_s$, when $g\Sigma(\tau) \approx 1$ the mode functions freeze-out and
become independent of time.  Notice that the largest $q$ modes have grown the
least, explaining why the
integral is dominated by $q \leq 10-20$.

For asymptotically large times, $g\Sigma$ is given by,
\begin{equation}\label{limfi}
g\Sigma(\infty) =g \; h^2 \; \int_0^{+\infty} {{dq}\over q } \;
\coth\left(\frac{ \omega_q}{2{\cal T}} \right) \; \left[ {{2h}\over {\pi }}
\mid c(q) \mid^2 - \, q \right] \;, \label{asinto}
\end{equation}
where only one term in the UV substraction survived in the $\tau =\infty $
limit.  For consistency, this integral must converge and be equal to $1$ as
given by the sum rule.  For this to be the case and to avoid the potential
infrared divergence in (\ref{asinto}), the coefficients $ c(q) $ must vanish at
$ q = 0 $. The mode functions are finite in the $ q \to 0 $ limit provided,
\begin{equation}
c(q) \buildrel{ q \to 0 }\over= {\cal C} \; q^{3/2} \; ,
\end{equation}
where $ {\cal C} $ is a constant.

The numerical analysis and figure (1.e) clearly show that the mode functions
remain finite as $q \rightarrow 0$, and the coefficient ${\cal C}$ can be read
off from these figures.  This is a remarkable result. It is well known that for
{\em free} massless minimally coupled fields in de Sitter space-time with
Bunch-Davies boundary conditions, the fluctuation contribution $\langle
\psi^2(\vec x, t) \rangle$ grows linearly in time as a consequence of the
logarithmic divergence in the
integrals\cite{linde2,vilford,vilenkin1}. However, in our case, although the
asymptotic mode functions are free, the coefficients that multiply the Bessel
functions of order $3/2$ have all the information of the interaction and
initial conditions and must lead to the consistency of the sum rule. Clearly
the sum rule and the initial conditions for the mode functions prevent the
coefficients $d(q)$ and $c(q)$ from describing the Bunch-Davies vacuum. These
coefficients are completely determined by the initial conditions and the
dynamics.  This is the reason why the fluctuation freezes at long times unlike
in the free case in which they grow linearly\cite{linde2,vilford,vilenkin1}.

It is easy to see from eqs.(\ref{finener})-(\ref{presfin}) and (\ref{uasi})
that the energy and presure vanish for $\tau \to \infty$.

Analogously, the two point correlation function can be computed in the late
time regime using the asymptotic results obtained above.  Inserting
eq. (\ref{asi}) for the mode functions in eq. (\ref{rhocorr}) yields the
asymptotic behavior:
\begin{equation}\label{Slargo}
S(\rho, \tau) \buildrel{\tau \to \infty}\over = { 1 \over {4\pi^2 \rho}} \;
\int_0^{\infty} {q \, dq} \;\sin q\rho \coth\left(\frac{ \omega_q}{2 {\cal
T}}\right) \; { {2 h^3} \over { \pi \, q^3}} \; \mid c(q) \mid^2 \; .
\end{equation}
The asymptotic behavior in time of the equal time correlation function is thus
solely a function of $ r $.  The large $ r $ behaviour of $ S(\vec{r}, +\infty)
$ is determined by the singularities of $ \mid c(k) \mid^2 $ in the complex $ k
$ plane. We find an exponential decrease,
\begin{equation}
S(\rho, +\infty) \buildrel{ \rho \to \infty }\over \simeq C \; {{e^{- \rho/ \xi
  } }\over \rho}\; ,
\end{equation}
where $ \rho = i / \xi $ is the pole nearest to the real axis and $ C $ is some
constant. Thus we see that the freeze-out of the mode functions leads to the 
freeze-out of the correlation length $\xi$. The result of the numerical
analysis is shown in figs. (1.g) and (2.b) which confirms
this behaviour and provides the
asymptotic value for $ \xi \approx 1/ h $. From these figures it is also clear
that the freeze-out time is given by the expansion time scale, $1/h$.  More
precisely, the numerical values for $ \xi $ can be accurately reproduced by the following formula obtained by a numerical fit
$$
h \xi \simeq 1.02 + 0.2 \ln h + 0.06 h + \ldots \; .
$$

This situation must be contrasted with that in Minkowski space-time
\cite{domains} where the correlation length grows as $\xi(\tau)
\approx \sqrt{\tau}$ during the stage of spinodal growth. Eventually, this
correlation length saturates to a fairly large value that is typically several
times larger than the zero temperature correlation 
length\cite{domains}.
We see that in the de Sitter case the domains are always horizon-sized.

\section{\bf Inflaton rolling down}

We now study the situation in which the inflaton zero mode rolls down the
potential hill and consider initial conditions such that $\eta(0) \neq 0 ;
\dot{\eta}(0) =0$.  In this case there will be two competing effects. One will
be the growth of spinodal fluctuations analyzed in the previous section, while
the other will be the rolling of the zero mode. Which effect will dominate the
dynamics is a matter of time scales. Before embarking on a numerical analysis
of these cases, it is illuminating to try to understand under what conditions
the dynamics will be driven by either the zero mode or the spinodal
fluctuations. We will analyze first the case of the large $N$ equations, and
compare later to the Hartree case.

The large $N$ equations are summarized by equations (\ref{dimlesslargenzero}) -
(\ref{largenincon}).  The important part of the mode equations that determine
the spinodal growth of long-wavelength modes is the term, $\eta^2 +g\Sigma$.
We can obtain an approximate estimate for the time scales on which the
contribution of the zero mode becomes significant as follows.  Let us consider
the situation in which $\eta(0) << 1$ and neglect the non-linear and
backreaction terms (the last two terms) in the equation of motion for the zero
mode (\ref{dimlesslargenzero}). At long times, but smaller than the time at
which either the quantum fluctuation or $\eta$ become of ${\cal O}(1)$, we
find,
\begin{equation}
\eta(\tau) \approx \eta(0) e^{(\nu - 3/2)h\tau}, \label{zeromodas}
\end{equation}
with $\nu$ given in eqn.(\ref{bessel}).  The estimate for the time scale for
the zero mode to be, $\eta_f \approx {\cal O}(1)$, is approximately given by,
\begin{equation}
\tau_{zm} \approx \frac{2}{(2\nu -3 )h}\ln\left[\frac{\eta_f}{\eta(0)}\right].
\label{zeromodetime}
\end{equation}
Comparing this time scale to the spinodal time scale given by
(\ref{spinotime}), for which quantum fluctuations grow to be of ${\cal O}(1)$,
we see that when,
\begin{equation}
\eta(0) << \lambda^{1/4}, \label{smallzero}
\end{equation}
the quantum fluctuations will grow to be ${\cal O}(1)$ much {\em earlier} than
the zero mode for $T_i > T_c$ (for $T_i=0$ the bound becomes $\eta(0) <<
\lambda^{1/2} $). In this case the dynamics will be driven completely
by the quantum fluctuations, as the zero mode will be rolling down the
potential hill very slowly and will not grow enough to compete with the quantum
fluctuations before the fluctuations grow to overcome the tree level terms in
the equations of motion.  In this case, as argued previously, the large $N$ and
Hartree approximations will be completely equivalent during the time scales of
interest.

On the other hand, if
\begin{equation}
\eta(0) >> \lambda^{1/4}, \label{bigzero}
\end{equation}
then the zero mode will roll and become ${\cal O}(1)$ {\em before} the
fluctuations have enough time to grow to ${\cal O}(1)$ ($\eta(0) >>
\lambda^{1/2} $ for $T_i=0$ ). In this case, the dynamics will be
dominated by the rolling of the zero mode and is mostly {\em classical}. The
quantum fluctuations remain perturbatively small throughout the inflationary
stage which will end when the velocity of the zero mode modifies the equation
of state to terminate de Sitter expansion.

For $\eta(0) \approx \lambda^{1/4}$ (or $\eta(0) \approx
\lambda^{1/2}$ for $T_i=0$), both the rolling of the zero mode {\em
and} the quantum fluctuations will give contributions of the same order to the
dynamics. In this case, the quantum fluctuations will be large for the
long-wavelength modes and the classical approximation to the inflationary
dynamics will not be accurate.

Since the scenario in which $\eta(0) >> \lambda^{1/4}$, in which the
dynamics is basically driven by the classical evolution of the zero mode has
received a great deal of attention in the literature, we will {\em not} focus
on this case, but instead analyze numerically the cases in which $\eta(0)
\neq 0$ but such that $\eta(0) \leq \lambda ^{1/4}$.

\subsection{\bf Numerical Analysis:}

We have evolved the set of equations of motion given by
(\ref{dimlesslargenzero}) and (\ref{dimlesslargenmodes}) numerically with
initial conditions (\ref{largenincon}) for the large $N$ case, and
(\ref{dimlesshartzero}) and (\ref{dimlesshartmodes}), with the corresponding
initial conditions (\ref{hartinconds}) on the mode functions for the Hartree
case.  The numerical code is the same as in the previous section with the same
relative errors.

{\bf Large N:}

Fig. (3.a,b) show $\eta(\tau)$ and $g\Sigma(\tau)$ vs. $\tau$ for the values
$\lambda=10^{-12}; \quad T_i/T_c=2; \quad \eta(0)= 10^{-5} ; \quad
\dot{\eta}(0)=0$.  Clearly the dynamics is dominated by the fluctuations; the
zero mode grows but is always negligibly small compared to $g\Sigma(\tau)$. The
time scale at which $g\Sigma(\tau)$ grows to be of order one is about the same
as in the case, $\eta(0)=0$, and all the behavior for the mode functions,
correlation length, energy density, pressure, etc. is similar to the case
analyzed in the previous section.

Asymptotically, we find that the sum rule,
\begin{equation}
-1+\eta^2(\infty)+g\Sigma(\infty)=0, \label{sumrulea}
\end{equation}
is satisfied to our numerical accuracy. This is the same as the situation in
Minkowski space-time\cite{frw,boyrehe}, and when $\eta \neq 0$, this sum rule
is nothing but the Ward identity associated with Goldstone's theorem. The
fluctuations are Goldstone bosons, minimally coupled, and the symmetry is
spontaneously broken with a very small expectation value for the order
parameter as can be read off from figure (3.a).  For $\tau > \tau_s$, the
dynamics freezes completely and the zero mode and the fluctuations achieve
their asymptotic values much in the same way as in the case $\eta =0$ studied
in the previous section.  Again, the correlation length becomes independent of
time with $\xi(\infty) \approx 1/h$ in a time scale given by $1/h$.

Because there is a damping term in the zero mode equation, it is reasonable to
assume that asymptotically there will be a solution with a constant value of
$\eta$.  Then the Ward-identity, $\eta(\infty)\left[
-1+\eta^2(\infty)+g\Sigma(\infty)\right]=0$, must be fulfilled. In the large
$N$ case, the {\em only} stationary solutions are i) $\eta =0 ; \;
g\Sigma(\infty)=1$, or, ii) $\eta(\infty) \neq 0 ; \;
-1+\eta^2(\infty)+g\Sigma(\infty)=0$. To have a consistent solution of the mode
functions, it must be that the effective mass term $\left[
-1+\eta^2(\infty)+g\Sigma(\infty)\right]$ vanishes asymptotically, leading to
the mode equations for massless, minimally coupled modes which are
asymptotically independent of time as shown in the previous section (see
eq.(\ref{uasi})). Furthermore, from fig. (3.b) it is clear that $g\Sigma(\tau)$
remains constant at long times, again unlike the case of free massless fields
with Bunch-Davies boundary conditions in which case the fluctuation grows
linearly in time\cite{linde2,vilford,vilenkin1}.

{\bf Hartree :} 

Figure (3.a) also shows the evolution of the zero mode in the large
$N$ and Hartree case. Although there is a quantitative difference in the
amplitude of the zero mode, in both cases it is extremely small and gives a
negligible contribution to the dynamics.  In the Hartree case, however, there
is no equivalent of the large $N$ sum rule; the {\em only} stationary solution
for $\eta \neq 0$ is, $\eta^2(\infty)=1 ; g\Sigma(\infty)=0$.  Such a solution
leads to mode equations with a {\em positive} mass term and mode functions that
vanish exponentially fast for $ \tau \to \infty $ for all momenta.  However,
whether the asymptotic behavior of the Hartree solution is achieved within the
interesting time scales is a matter of initial conditions. For example in
fig. (3.a) the initial condition is such that the time scale for growth of the
quantum fluctuations is much shorter than the time scale for which the
amplitude of the zero mode grows large and the non-linearities become important.
In the large $N$ case the sum rule is satisfied with a large value of the
quantum fluctuations. In the Hartree case the equivalent sum rule
$-1+3\eta^2_H+3g\Sigma_H=0$ is satisfied for a very small $\eta_H$ and a
$g\Sigma_H \approx 1/3$.  The modes become effectively massless and they stop
growing.

The equation for the zero mode (see eqn. (\ref{dimlesshartzero})) still has an
uncancelled piece of the non-linearity, $-2\eta^2$; however the derivatives and
the amplitude of $\eta$ are all extremely small and though the zero mode
still evolves in time, it does so extremely slowly.  In fact the Hartree curve
in fig. (3.a) has an extremely small positive slope asymptotically, and while
$\eta_H$ grows very slowly, $g\Sigma_H$ diminishes at the same rate. In the
case shown in fig. (3.a), we find numerically that $\dot{\eta}_H / \eta_H
\approx 10^{-7}$ at $\tau= 150$.  Before this time most of the interesting
dynamics that can be captured with a fixed de Sitter background had already
taken place, and the backreaction of the fluctuations on the metric becomes
substantial requiring an analysis that treats the scale factor dynamically. 

The conclusion of our analysis is that in the region of initial conditions for
which the quantum fluctuations dominate the dynamics, that is for $\eta(0) <<
\lambda^{1/4}$, both large $N$ and Hartree give the same answer on
 the relevant time scales. 
The figures for ${\cal H}(\tau) / {\cal H}(0)$ are
numerically indistinguishable from the case of figures (1).

For comparison, we show in figure (4) both cases for the zero mode, for the
same values of the parameters as in figures (1,3) but with the initial
condition $\eta(0)=10^{-3}\quad ; \dot{\eta}(0)=0$. This is a borderline case
in which the time scales for the evolution of the zero mode and quantum
fluctuations are of the same order and there is no clear separation of time
scales between these two competing terms in the evolution equations.

We see that in the large $N$ case the zero mode rolls to a final amplitude
which is ${\cal O}(1)$ and of the same order as $g\Sigma(\infty)$ and the sum
rule is satisfied.  However, the Hartree case clearly shows the asymptotics
analyzed above with $\eta_H(\infty)=1 ; g\Sigma_H(\infty)=0$.

This particular borderline case is certainly not generic and would imply some
fine tuning of initial conditions. Finally the case in which $\eta(0) >>
\lambda^{1/4}$ (or $\lambda^{1/2}$ for $T_i=0$) is basically classical in that
the dynamics is completely given by the classical rolling of the zero mode and the fluctuations are always perturbatively small.

\section{Discussion and Conclusions}

We have identified analytically and numerically two distinct regimes for the
dynamics determined by the initial condition on the expectation value of the
zero mode of the inflaton.

\begin{enumerate}
\item{When $\eta(0) << \lambda^{1/4}$ (or $\lambda^{1/2}$ for $T_i=0$), the
dynamics is driven by quantum (and thermal) fluctuations. Spinodal
instabilities grow and eventually compete with tree level terms at a time
scale, $\tau_s \geq -3h\ln[\lambda]/2$.  The growth of spinodal fluctuations
translates into the growth of spatially correlated domains which attain a
maximum correlation length (domain size) of the order of the horizon.  For very
weak coupling and $h \geq 1$ this time scale can easily accomodate enough
e-folds for inflation to solve the flatness and horizon problems. The quantum
fluctuations modify the equation of state dramatically and at this time scale
can modify the dynamics of the scale factor and provide a means for a graceful
exit to the inflationary stage without slow-roll.

This non-perturbative description of the non-equilibrium effects in this
regime in which quantum (and thermal) fluctuations are most important
is borne out by both the large N and Hartree approximations. Thus
our analysis provides a reliable understanding of the relevant 
non-perturbative, non-equilibrium effects of the fluctuations that have
not been revealed before in this setting.

These initial conditions are rather natural if the de Sitter era arises during
a phase transition from a radiation dominated high temperature phase in local
thermodynamic equilibrium, in which the order parameter and its time derivative
vanish.}

\item{When $\eta(0) >> \lambda^{1/4}$ (or $\lambda^{1/2}$ for $T_i=0$), the
dynamics is driven solely by the classical evolution of the inflaton zero
mode. The quantum and thermal fluctuations are always perturbatively small
(after renormalization), and their contribution to the dynamics is negligible
for weak couplings. The de Sitter era will end when the kinetic contribution to
the energy becomes of the same order as the `vacuum' term. This is the realm of
the slow-roll analysis whose characteristics and consequences have been
analyzed in the literature at length. These initial conditions, however,
necessarily imply some initial state either with a biasing field that favors a
non-zero initial expectation value, or that in the radiation dominated stage,
prior to the phase transition, the state was strongly out of equilibrium with
an expectation value of the zero mode different from zero. Although such a
state cannot be ruled out and would naturally arise in chaotic scenarios, the
description of the phase transition in this case requires further input on the
nature of the state prior to the phase transition.}
\end{enumerate}

We have learned from this work that non-equilibrium effects can alter scalar
field dynamics in a dramatic way, under well specified and physically
reasonable conditions. In particular, this results bring up the tantalizing
possibility that when the scale factor is coupled to the inflaton dynamically,
in a full backreaction treatment, some of the standard results concerning the
time evolution of the inflaton could be modified in unexpected ways. The fact
that we found massless fields whose fluctuations did {\em not} grow linearly is
an extremely interesting result, especially in light of how the quantum
fluctuations become density perturbations. We are currently working on the
formalism that allows us to couple gravity dynamically in a non-equilibrium
way. Furthermore, it would be interesting to extend the program of
reconstruction of the inflaton effective potential (see\cite{reconst}) to
include the possibility of the dynamics being driven by spinodal fluctuations
and not by ``slow roll'' of the inflaton zero mode.

The results presented here raise some further interesting questions: how do
these fluctuations contribute to the spectrum of primordial scalar density
perturbations? How should the approach to cosmological perturbations based on
slow-roll be modified to the case studied here in which spinodal fluctuations,
{\em not} slow-roll drive the dynamics?  We are currently studying these issues
within the consistent non-perturbative, non-equilibrium program presented in
this article.

\acknowledgements 
D. B. would like to thank the N.S.F for partial support through the grant
awards: PHY-9302534 and INT-9216755, the Pittsburgh Supercomputer Center for
grant award No: PHY950011P and LPTHE for warm hospitality.  R. H. and
D. C. were supported by DOE grant DE-FG02-91-ER40682.

\newpage

\centerline{\bf Figure Captions:}

{\bf Fig. 1a:} $g\Sigma(\tau)$ vs. $\tau$ for $\eta(0)=\dot{\eta}(0)=0; \quad
\lambda=10^{-12}; \quad r=2; \quad h=2$.

\vspace{2mm}

{\bf Fig. 1b:} $\lambda\left[p(\tau)+\varepsilon(\tau)\right]/2m^4_r$
vs. $\tau$ for the same values of parameters as in fig. 1a.

\vspace{2mm}

{\bf Fig. 1c:} ${\cal H}(\tau) / {\cal H}(0)$ vs $\tau$ for the same parameters
as in fig. 1a.

\vspace{2mm}

{\bf Fig. 1d:} $\dot{\cal H}(\tau) / {\cal H}^2(\tau)$ vs $\tau$ for the same
values as in fig. 1a.

\vspace{2mm}

{\bf Fig. 1e:} $g{\cal S}(q,\tau)$ vs. $q$ for $\tau= 60$. Same parameters as
in fig. 1a.

\vspace{2mm} 

{\bf Fig. 1f:} $S(\rho,\tau) / S(0,\tau)$ vs. $\rho$ for $\tau \ge 2$. Same
parameters as in fig. 1a.

\vspace{2mm}

{\bf Fig. 1g:} $\xi(\tau)$ vs $\tau$. Same parameters as in fig. 1a.

\vspace{2mm}

{\bf Fig. 1h:} $\ln\left[|f_q(\tau)|^2\right]$ vs. $\tau$ for $q=0,4,10$. Same
parameters as in fig. 1a.

\vspace{2mm}

{\bf Fig. 2a:} $g\Sigma(\tau)$ vs. $\tau$ for $\eta(0)=\dot{\eta}(0)=0; \quad
; \lambda=10^{-12} \quad r=2; \quad h=0.1$.

\vspace{2mm}

{\bf Fig. 2b:} $\xi(\tau)$ vs $\tau$. Same parameters as in fig. 2a.

\vspace{2mm}

{\bf Fig. 3a:} $\eta(\tau)$ vs. $\tau$ for $\lambda=10^{-12}; \quad r=2; \quad
h=2; \quad \eta(0)=10^{-5};\quad \dot{\eta}(0)=0$ for large $N$ (solid curve)
and Hartree (dashed curve).

\vspace{2mm}

{\bf Fig. 3b:} $g\Sigma(\tau)$ vs. $\tau$ for the same values of the
parameters
as in fig. 3a, for large $N$ (solid curve) and Hartree (dashed curve).

\vspace{2mm}

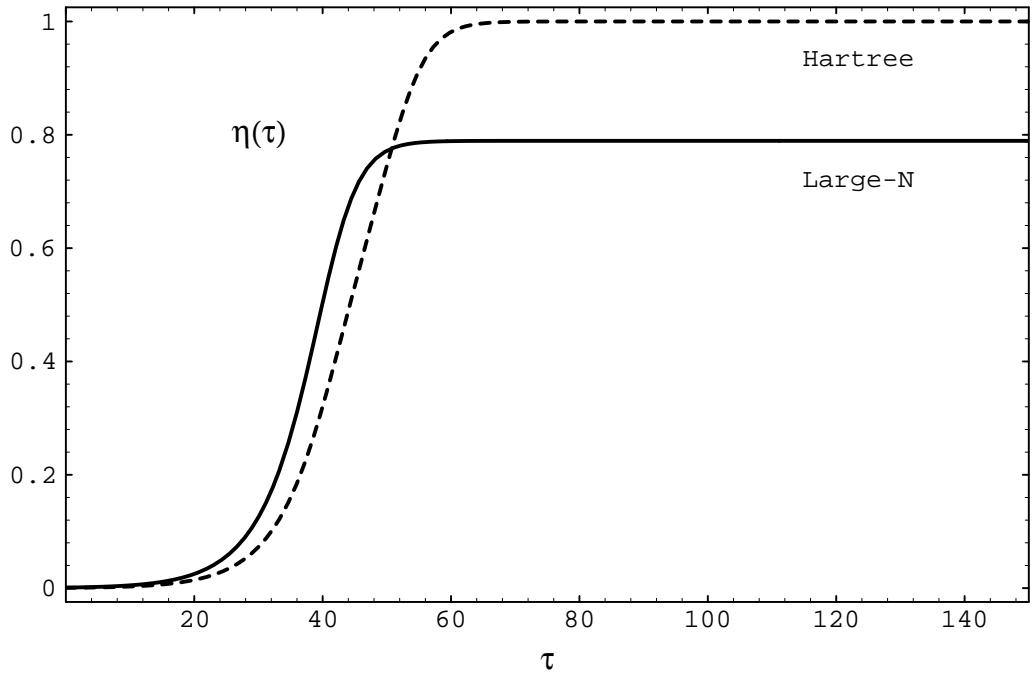
{\bf Fig. 4:} Comparison of zero mode dynamics for large $N$ (solid curve) and
Hartree (dashed curve) for $\eta(0)=10^{-3} ; \quad \dot{\eta}(0)=0$ and all
other parameters as in figures (1,2).

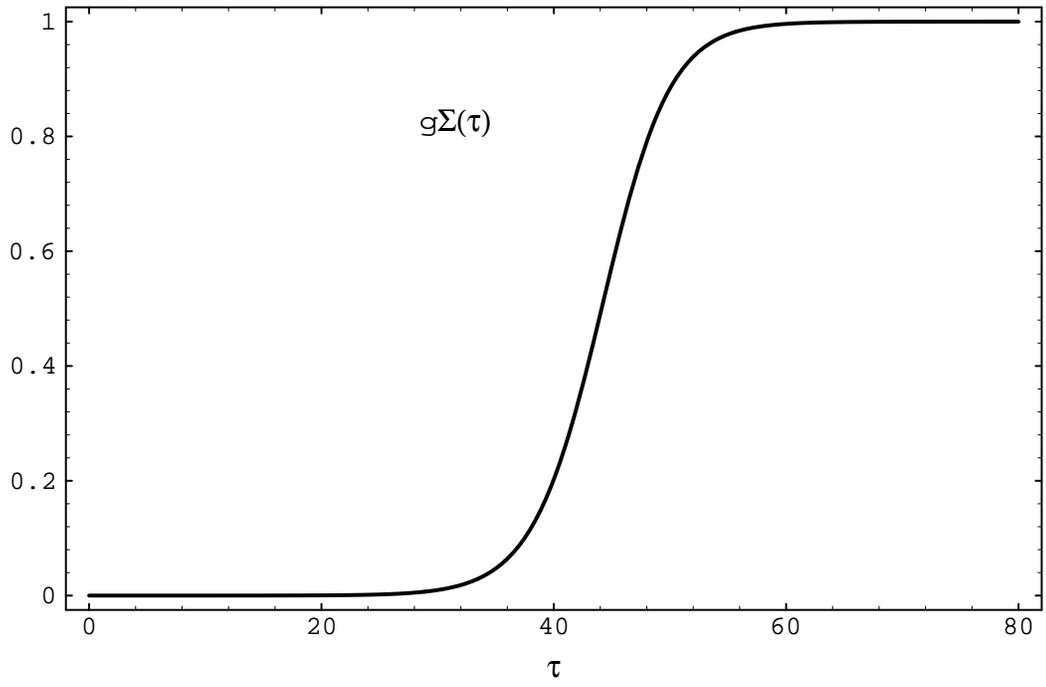
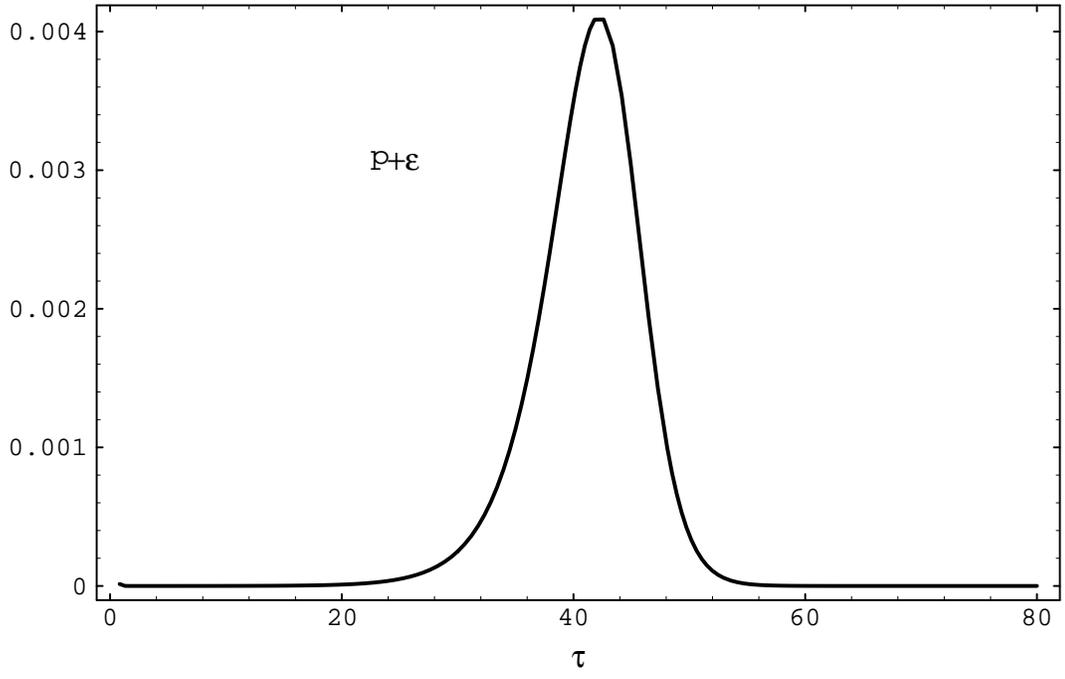
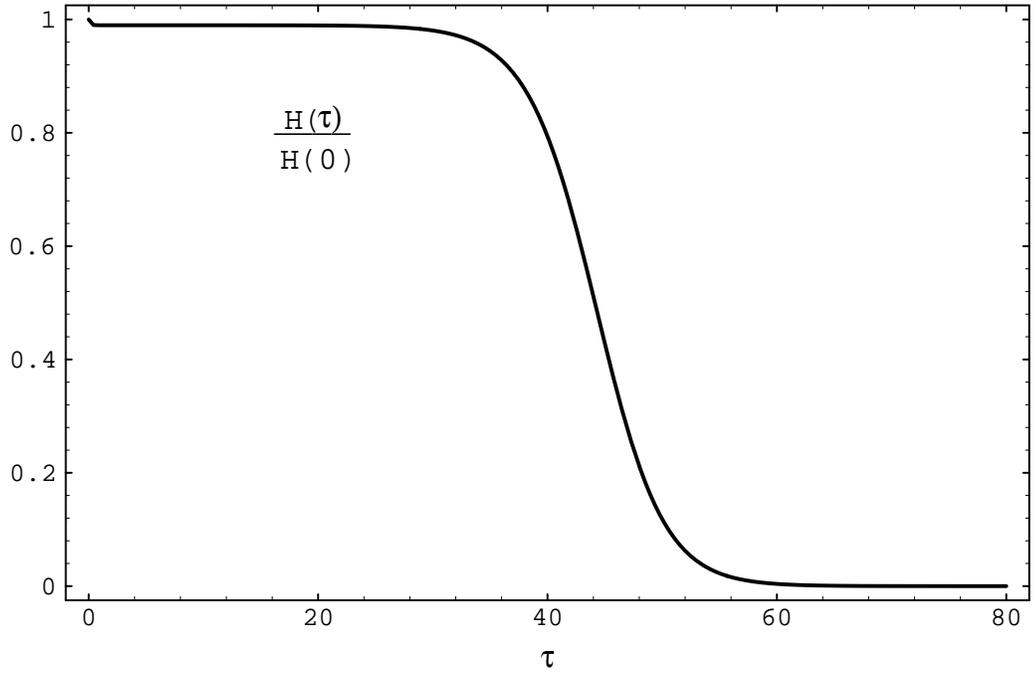
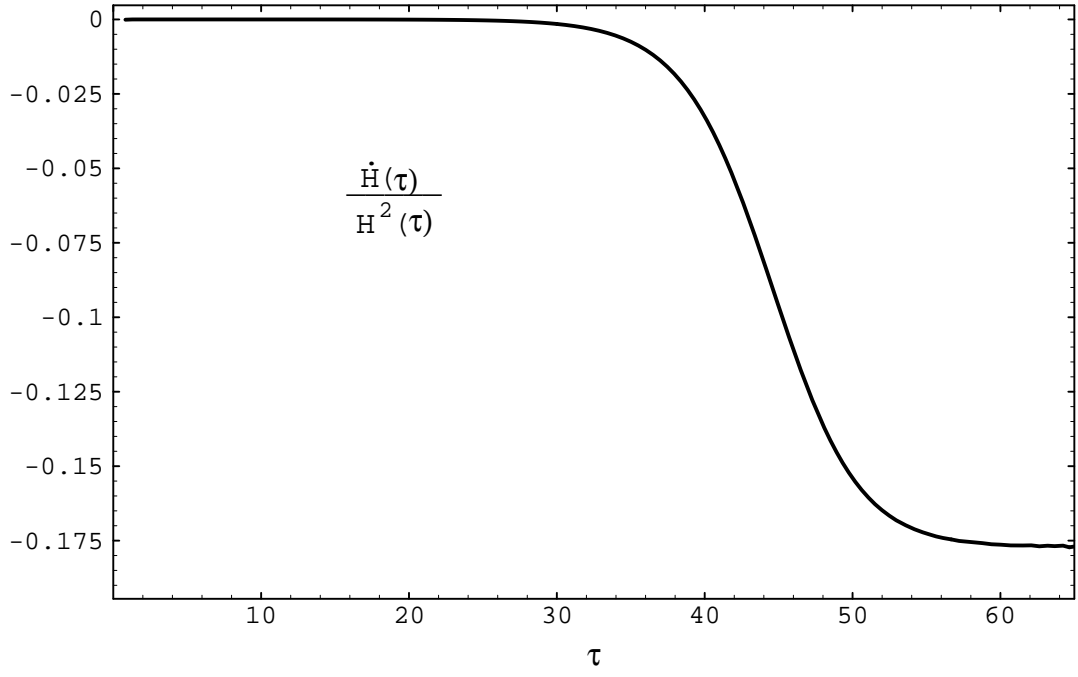
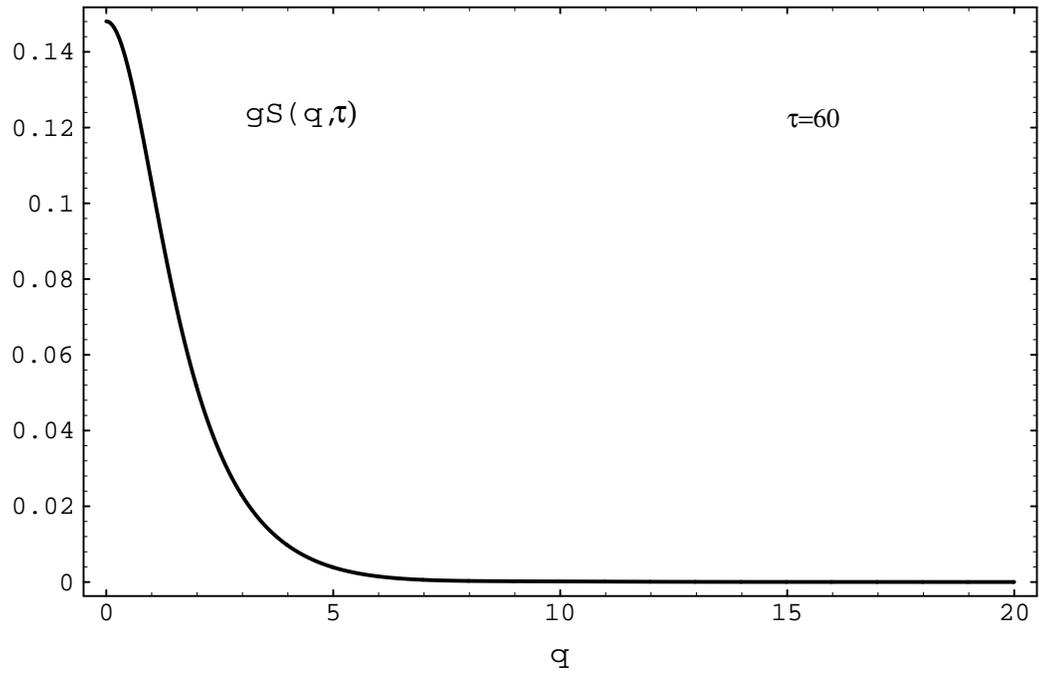
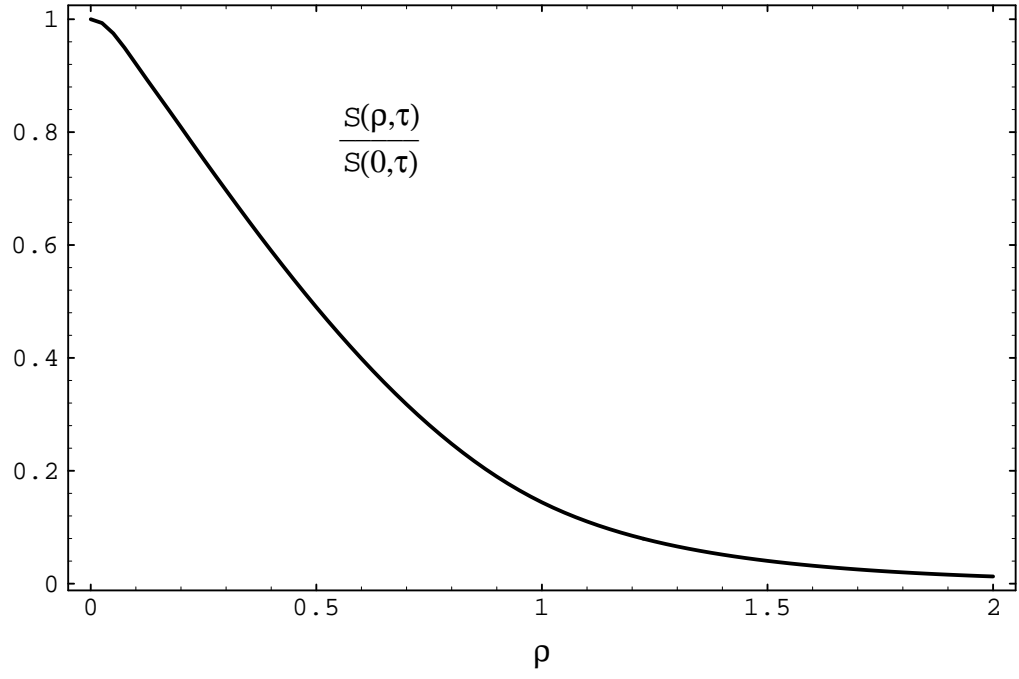
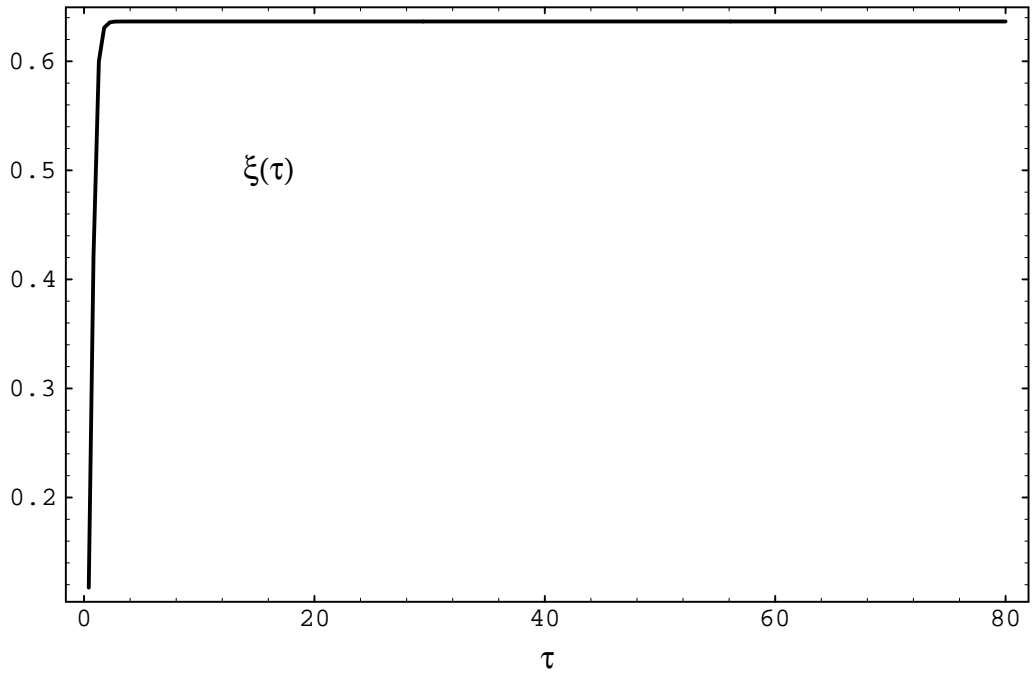
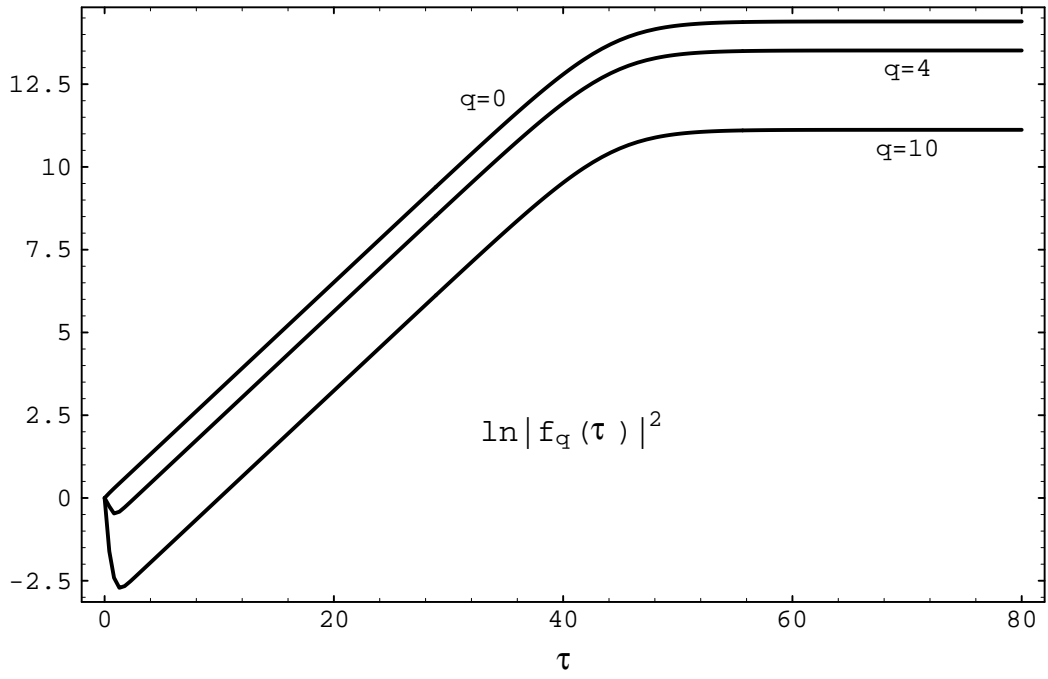
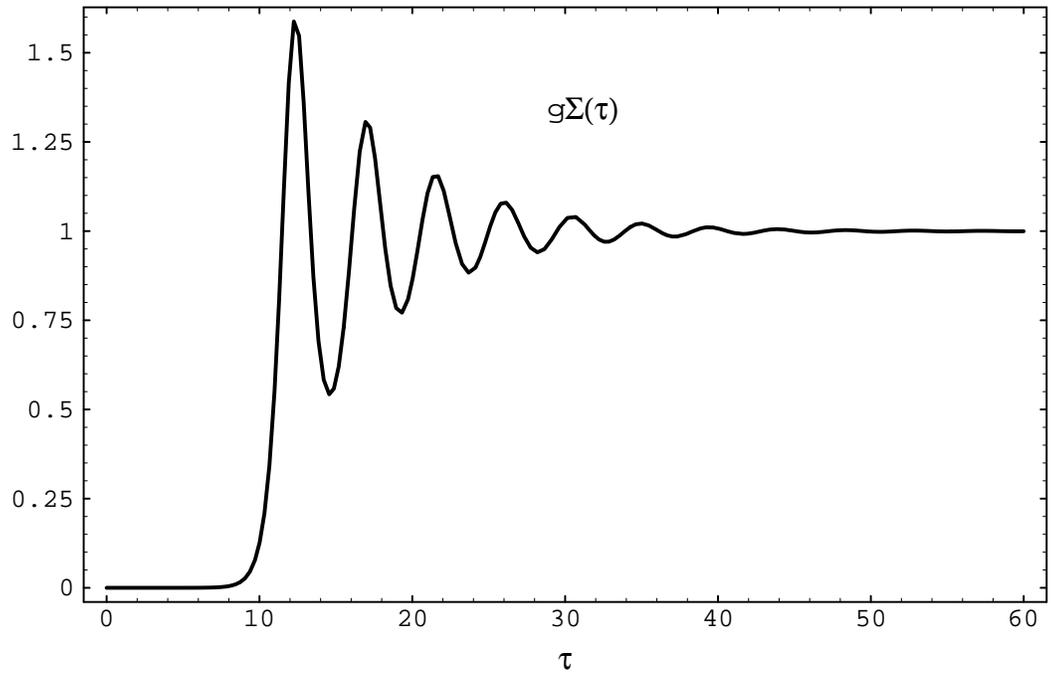
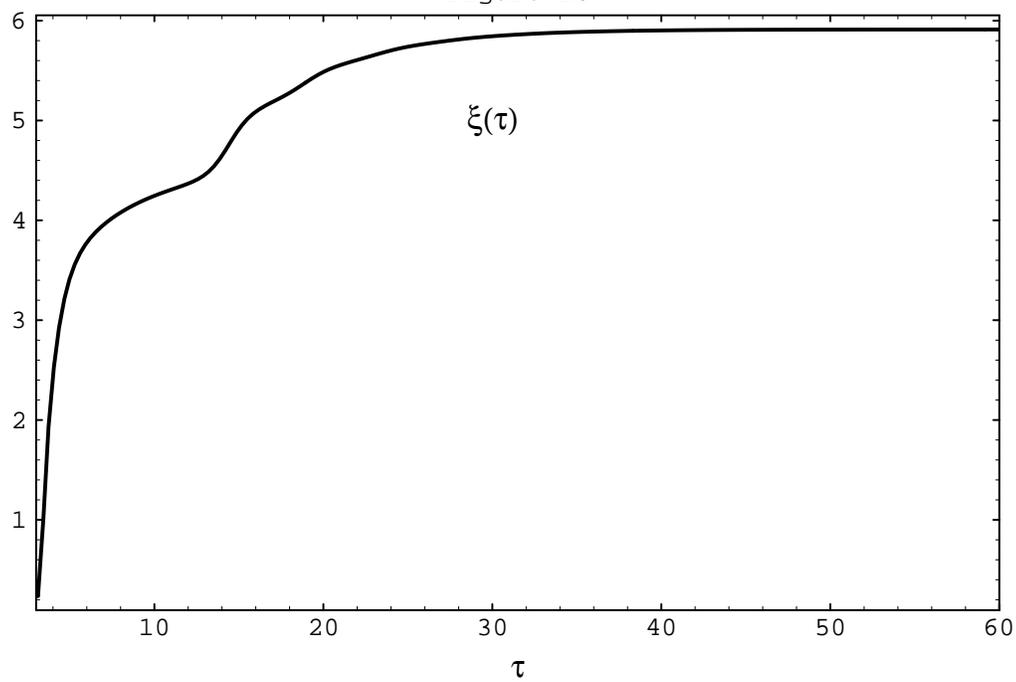
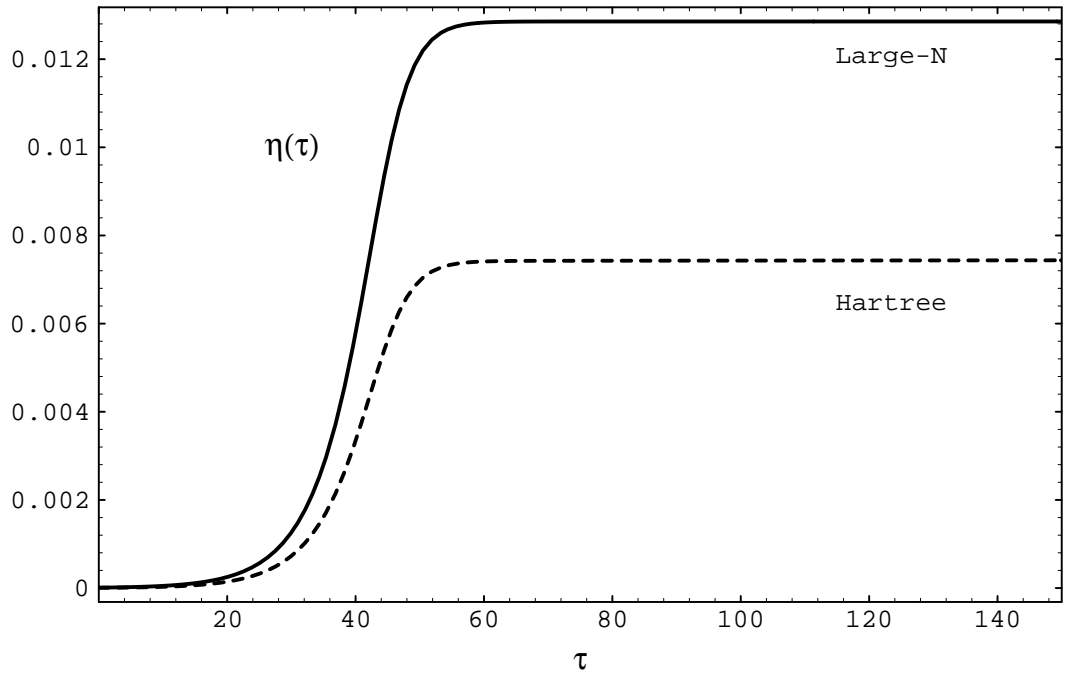
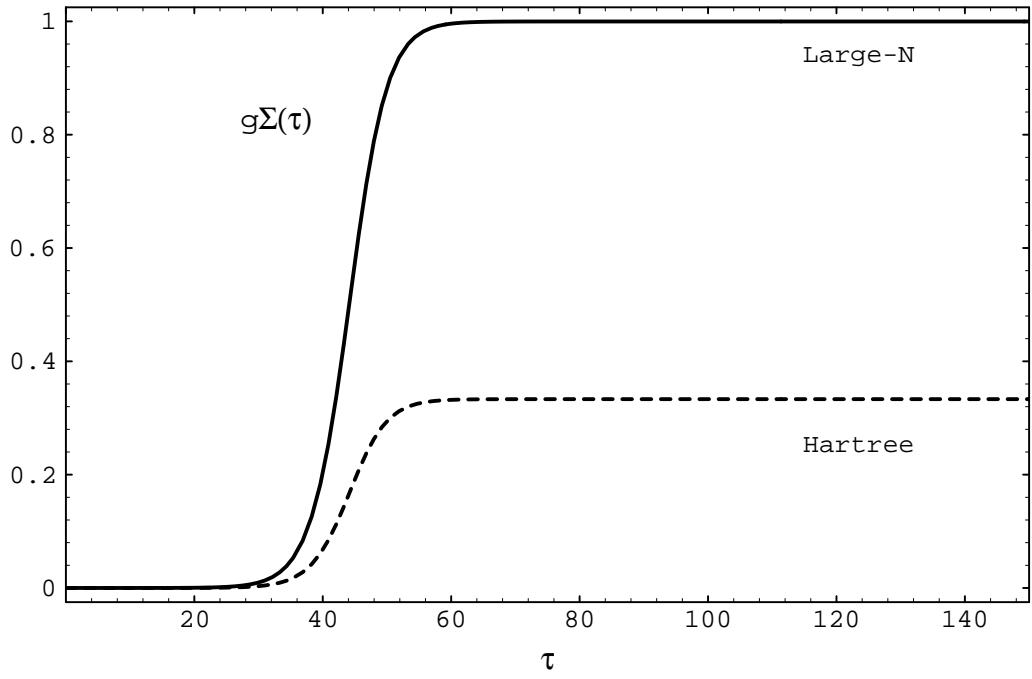
\end{document}